\documentclass[aps]{revtex4-1}

\usepackage{dcolumn}
\usepackage{amsbsy}
\usepackage{amsmath}

\usepackage{amssymb}
\usepackage{braket}
\usepackage{color}
\usepackage{fullpage}
\usepackage{graphicx}
\usepackage{bm}

\usepackage{subfigure}
\usepackage[hidelinks]{hyperref}
\usepackage{epstopdf}

\usepackage{comment}
\graphicspath{{./Figures/}}

\newcommand{\bK}{\bm{K}}
\newcommand{\bM}{\bm{M}}

\newcommand{\cD}{\mathcal{D}}

\newcommand{\boldeta}{\boldsymbol{\eta}}

\begin{document}
\title{Topological bands and localized vibration modes in quasiperiodic beams}
\author{Raj Kumar Pal$^{a}$, Matheus I.N. Rosa$^{b}$ and Massimo Ruzzene$^{b,c}$ \\
\small $^a$ Department of Mechanical and Civil Engineering, California Institute of Technology, Pasadena CA, 91125 \\
\small $^b$ School of Mechanical Engineering, Georgia Institute of Technology, Atlanta GA 30332 \\
\small $^c$ School of Aerospace Engineering, Georgia Institute of Technology, Atlanta GA 30332}

\begin{abstract}
We investigate a family of quasiperiodic continuous elastic beams, the topological properties of their vibrational spectra, and their relation to the existence of localized modes. We specifically consider beams featuring arrays of ground springs at locations determined by projecting from a circle onto an underlying periodic system. A family of periodic and quasiperiodic structures is obtained by smoothly varying a parameter defining such projection. Numerical simulations show the existence of vibration modes that first localize at a boundary, and then migrate into the bulk as the projection parameter is varied. Explicit expressions predicting the change in the density of states of the bulk define topological invariants that quantify the number of modes spanning a gap of a finite structure. We further demonstrate how modulating the phase of the ground springs distribution causes the topological states to undergo an edge-to-edge transition. The considered configurations and topological studies provide a framework for inducing localized modes in continuous elastic structural components through globally spanning, deterministic perturbations of periodic patterns defined by the considered projection operations.
\end{abstract}

\maketitle

\section{Introduction}

A notable feature of disordered (or aperiodic) media is their ability to support localized eigenmodes, in contrast to Bloch modes which span entire domains in periodic media~\cite{anderson1958absence}. Numerous studies have investigated localized states in a variety of physical systems, including optics and elastic media with disorder~\cite{hu2008localization}, random inclusions~\cite{sheng1990scattering,han2008wave} and nonlinear interactions~\cite{sievers1988intrinsic,campbell2004localizing,page1990asymptotic}. In spite of abundant literature on the existence of localized states in disordered media, there are still fundamental questions that remain unaddressed regarding the nature of these modes, such as their robustness to defects and imperfections, and the effects of finite size, of interactions with other fields, and of nonlinearities~\cite{fishman2012nonlinear}.

As a particular class of non-periodic media, quasiperiodic (QP) media are characterized by deterministic patterns that retain long range order and may exhibit symmetries that are forbidden in periodic media, such as rotational symmetries other than $2,3,4$ and $6$-fold~\cite{janot2012quasicrystals}. For these reasons, QP media exhibit unique properties that have been the subject of extensive investigations in diverse areas of physics, including electronics~\cite{steinhardt1987physics}, electromagnetics~\cite{man2005experimental} and elasticity~\cite{liu2004governing}. One of the hallmarks of QP media is in the fractal spectra, which was originally observed for electrons in the presence of magnetic fields exhibiting the well-known Hofstadter's butterfly spectrum~\cite{hofstadter1976energy}. In the context of localization, the transition from globally spanning to localized modes originally introduced by Aubrey and Andr{\'e}~\cite{aubry1980analyticity} has been observed in different QP systems, such as photonic lattices~\cite{lahini2009observation} and elastic granular chains~\cite{martinez2018quasiperiodic}. These localized modes are of interest to physical phenomena like disorder enhanced transport~\cite{segev2013anderson}, or one-way reflection in acoustic waveguides~\cite{zhu2018simultaneous}.    

The deterministic nature of QP media allows for investigations of their properties through a framework that is rarely available for other type of disordered systems, such as those characterized by random inclusions. In particular, a recent line of work regards QP lattices as projections of higher dimensional manifolds onto lower dimensional lattices~\cite{kraus2016quasiperiodicity,ozawa2016synthetic}, and such notion is used to explore topological properties of higher dimensional periodic systems~\cite{kitaev2009periodic}. Notable examples include the works of Zilberberg and coworkers, who used photonic waveguides to achieve adiabatic pumping of waves between opposite ends of one-dimensional (1D) QP lattices~\cite{kraus2012topological,verbin2013observation}. With further studies, the authors experimentally demonstrated a dynamically generated four-dimensional (4D) quantum Hall system through two-dimensional (2D) periodic lattices~\cite{zilberberg2018photonic}. In the classic mechanics realm, Prodan and coworkers~\cite{apigo2018topological} have recently shown localized modes at the boundary of a QP chain of magnetic spinners, and experimentally demonstrated topological boundary and interface modes in acoustic waveguides with QP patterning of the walls~\cite{apigo2019observation}. Although open questions still remain regarding how eigenmodes may localize in any region of the domain, these studies provide insight into modes that are localized at edges or interfaces, and suggest new methodologies for wave localization and transport based on higher dimensional topological properties.
 
Motivated by these contributions, we here investigate how localized modes arise in continuous elastic media with QP stiffness modulations. Such modulations are introduced through arrays of ground springs embedded in structural beams undergoing transverse vibrations (Fig.~\ref{Fig: config}). The locations of the springs are determined by projecting from periodic array of circles. This pattern-generating procedure identifies families of structures ranging from periodic to QP, that are obtained through the smooth variation of the parameters defining the projection. This study contributes to the recent investigations of the dynamic response of QP continuous elastic media~\cite{gei2010wave,morini2018waves,chen2008elastic}. For example, Gei and coworkers~\cite{gei2010wave,morini2018waves} have analyzed the self-similar and invariant nature of stop and pass bands in beams and rods embedded with a QP array of supports and springs, while Wang and coworkers~\cite{chen2008elastic} demonstrated localization in plates with a QP array of inclusions. While prior studies are notable in observing stop bands and localized modes, a formal treatment of their topological properties and connection to the existence of localized modes is currently missing. Furthermore, there is limited understanding of how these modes can be induced in desired regions of continuous elastic media. Towards bridging these gaps, we here investigate the spectral properties of QP beams as exemplary continuous structural components. Specifically, this study shows that the topological properties of their vibrational spectra are related to the presence of localized modes, and that smooth variation of the projection parameters leads these modes through transitions, whereby they are globally spanning in periodic structures, become localized at boundaries, and then migrate to the bulk, or interior, of structural domains. Explicit expressions for the density of states of the bulk spectrum define topological invariants that provide the number of edge modes spanning a gap for a finite structure. We further demonstrate that phase modulations of the QP patterns drive the topological states through transitions as they merge with the bulk and change their localization edge. 

The outline of this paper is as follows: following this introduction, Section~\ref{theorySec} presents a description of the QP beams, the governing equations and the approximate solution procedure. Section~\ref{spectraSec} presents an analysis of the frequency spectra of finite and infinite structures and illustrates topologically protected modes in finite beams and their transitions driven by phase modulations. Finally, Section~\ref{concSec} summarizes the main results of the work and outlines future research directions. 

\section{Quasiperiodic configurations and analytical framework}\label{theorySec}
\begin{figure}
	\centering
		\includegraphics[width=0.48\textwidth]{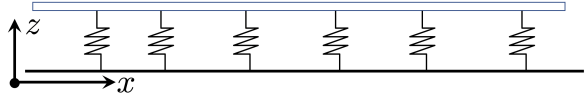}
	\caption{Schematic of beam with ground springs. } 
	\label{Fig: config}
\end{figure}

The beams considered here are equipped with arrays of elastic springs that connect them to the ground at locations defined by QP 1D patterns (Fig.~\ref{Fig: config}). These continuous elastic structures can be analyzed within a general framework that allows the investigation of the dynamic behavior of interest, which includes natural frequencies and mode shapes, and their dependence on the location of the inclusions. 

\subsection{Periodic and quasiperiodic patterns}\label{patternsSec}
\begin{figure}[b!]
	\centering
	\subfigure[]{
		\includegraphics[width=0.7\textwidth]{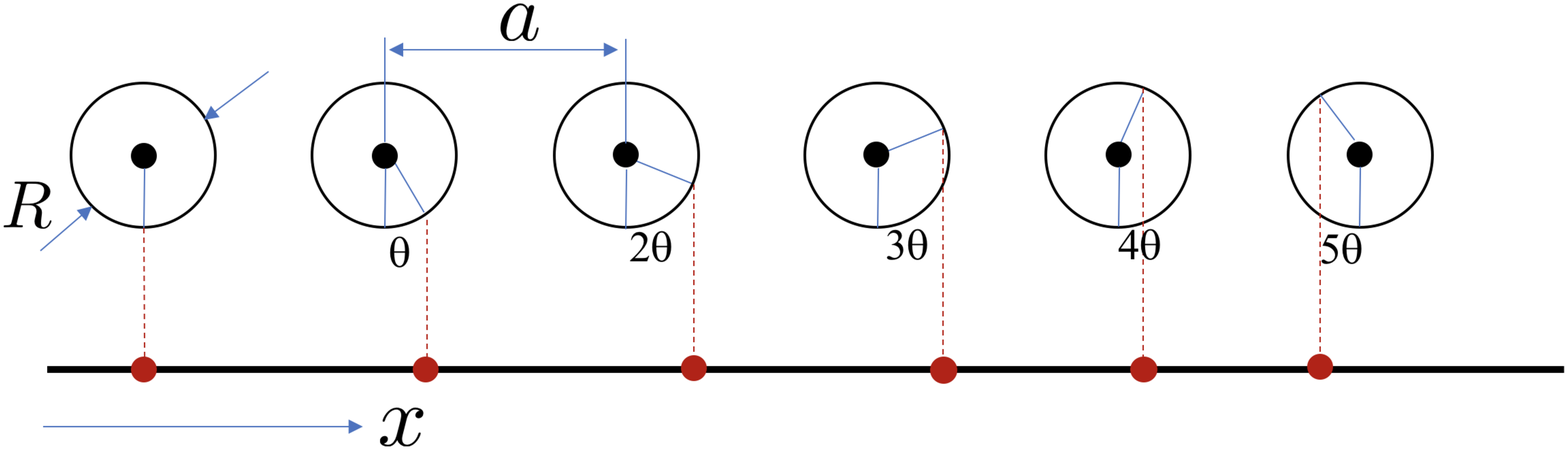}\label{QP_beam_scheme}}\\
	\subfigure[]{
		\includegraphics[width=0.7\textwidth]{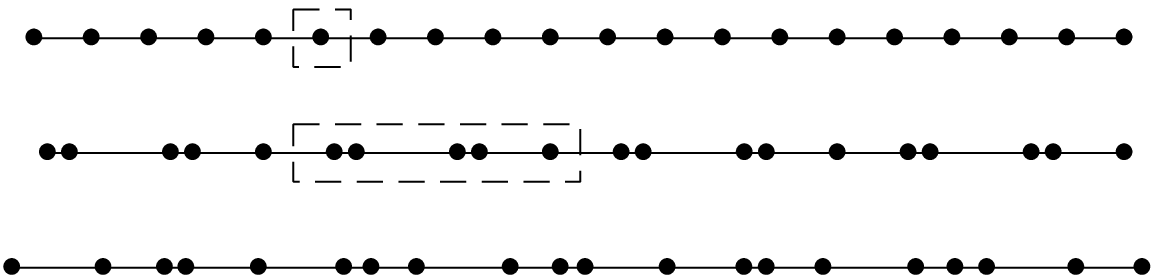}\label{QP_beam_scheme3}
	}
	\caption{(a) Generation of a 1D patterns through projections from a circle ~\cite{apigo2018topological}. (b) Periodic and QP sequences corresponding to three different values of the parameter $\theta$:  $\theta=0$ and $\theta=0.4=2/5$ define two periodic sequences (top and middle) of unit cell highlighted by the dashed rectangle, while $\theta=1/\sqrt{2}$ identifies a QP pattern with no spatial translational symmetry.} 
	\label{Fig: QP 1D}
\end{figure}

The ground springs are located according to patterns generated through the projection approach described in~\cite{apigo2018topological}. Specifically, the springs location $x_s$ are obtained from an underlying periodic lattice of period $a$, and by projecting from a manifold, which here is a circle of radius $R<a/2$ centered at the periodic sites $s a$, $s \in \mathbb{Z}$ (Fig.~\ref{QP_beam_scheme}). Thus, the location of spring $s$ is given by
\begin{equation}\label{eq: beamQP}
x_s = s a + R \sin 2\pi s \theta,
\end{equation}
where $ \theta \in [0 , 1]$ is the \emph{projection parameter} that governs the wavelength of the modulation with respect to the reference spacing $a$. Rational and irrational $\theta$ values respectively define periodic and QP patterns, as illustrated in Fig.~\ref{QP_beam_scheme3} which shows examples for $\theta = 0$, $0.4$, $1/\sqrt{2}$. The pattern with $\theta=0$ has periodicity $a$ with the unit cell indicated by the dashed rectangle. Rational $\theta= p/q$, with $p$ and $q$ being coprime integers, result in periodic patterns with unit cells of length $qa$. For example, the dashed rectangle in the middle pattern of Fig.~\ref{QP_beam_scheme3} shows a periodic unit cell for the pattern with $\theta=0.4=2/5$, which comprises of $q=5$ locations. In contrast, no such periodic unit cell exists for the $\theta=1/\sqrt{2}$ QP pattern or any other irrational $\theta$ value. 

\subsection{Governing equations}\label{govEqnSec}
The equations governing the transverse motion of beams (Fig.~\ref{Fig: config}) can be generally expressed as~\cite{meirovitch1975elements}
\begin{equation}\label{eq: eom general}
\mathcal{L}[w (x, t)] + \rho A \frac{\partial^2 w(x,t)}{\partial t^2}=0, \quad x \in \cD
\end{equation}
where $w(x,t)$ is the displacement in the transverse $z$ direction at point $x$ and time $t$, $\rho$ is the mass density, $A$ is the cross sectional area, $\cD$ is the domain of the beam and $\mathcal{L}[\cdot]:H^2(\cD)\to L^2(\cD)$ is a linear self-adjoint spatial differential operator.
Note that $w\in H^2(\cD)$ has continuous first order derivatives, is twice differentiable almost everywhere in $\cD$ and its higher derivatives are defined in a weak or distributional sense. For 1D beams, the differential operator  $\mathcal{L}[\cdot]$ formulated according to Euler-Bernoulli theory is given by
\begin{equation}\label{eq: operator beam}
\mathcal{L} =EI\dfrac{\partial^4}{\partial x^4} + k_g \sum_{s=1}^S  \delta(x-x_s),
\end{equation}
Here $E$ denotes the material Young's modulus and $I$ is the second moment of area. Also, $\delta$ is the Dirac delta function, while $k_g$ denotes the ground spring constant. Under the assumption of harmonic motion at frequency $\omega$, Eqn.~\eqref{eq: eom general} can be written as an eigenproblem
\begin{equation}\label{eq: eom general frequency}
\mathcal{L}[w (x,t)]=\rho A\omega^2w(x,t)
\end{equation}

\subsection{Approximate solution using Galerkin's method}
The vibration characteristics of QP beams are investigated by employing
Galerkin's method~\cite{ern2013theory} applied to finite domains $\cD$. An approximate solution of the eigenproblem given by Eqn.\eqref{eq: eom general frequency} is sought in terms of a set of $N$  comparison functions $\phi_n$ and generalized coordinates $\eta_n$ as
\begin{equation}\label{eq: Galerkin}
w(x)=\sum_{n} \phi_n(x) \eta_n .
\end{equation}
We seek the weak solution of Eqn.~\eqref{eq: eom general frequency} with comparison functions 
in the subspace spanned by the same comparison functions. In particular, $\eta_i$ are determined by imposing that the weighted error $\epsilon$ integrated over the domain $\cD$ is zero, which may be written as
\begin{equation}
\int_{\cD} \epsilon \phi_n({x}) d \cD = \int_{\cD} ({\mathcal{L}}[w] - \rho A\omega^2 w) \phi_n({x}) d \cD = 0 . 
\end{equation}
This condition leads to a set of algebraic equations in the form 
\begin{equation}
\sum_{i} (k_{ij}-\omega^2 m_{ij}) \eta_i = 0, \,\,\, i,j=1,..,N
\end{equation}
where
\begin{equation}\label{eqn: Kstiff}
k_{ij} =\int_{\cD} {\mathcal{L}}[\phi_i({x})] \phi_j({x}) d \cD, \,\,\, m_{ij} =\rho A\int_{\cD} \phi_i({x}) \phi_j({x}) d \cD .
\end{equation}
The resulting discretized eigenvalue problem in matrix form is
\begin{equation}\label{eq: eig discrete}
\bK \boldeta=\omega^2 \bM \boldeta , 
\end{equation}
where $\bK,\bM \in \mathbb{R}^{N \times N}$ and $\boldeta \in \mathbb{R}^N$. 

\section{Spectral Properties of Bulk and Finite Domains}\label{spectraSec}
The vibration characteristics of QP beams are investigated by employing
Galerkin's method~\cite{ern2013theory} applied to finite domains $\cD$. The number of unit cells included in the computations depends on the objective of the investigations, which is two-fold. We first investigate structures that are defined by sufficiently large domains and are representative of the bulk properties of infinite periodic and QP domains~\citep{apigo2018topological}. This analysis reveals a fractal frequency spectrum where the associated integrated density of states (IDS) define topological invariants that predict the existence of localized modes. The existence of such modes in structures of finite extent is then evaluated by considering proper termination of the domains through prescribed sets of boundary conditions. The results shown herein were obtained by using a fixed radius $R=0.4a$ defining the QP patterns, while a non-dimensional frequency is defined as $\Omega=\omega/\omega_0$, with $\omega_0=\sqrt{(EI)/(\rho Aa^4)}$. Also, we define a non-dimensional ground spring stiffness $\gamma_g=(a^3k_g)/(EI)$ that corresponds to the ratio between the ground spring stiffness $k_g$ and the equivalent static stiffness $EI/(a^3)$ of a beam segment with length $a$.

\subsection{The bulk spectrum and its topology}

We consider a beam with a large number $S$ of springs with periodic boundary conditions imposed on both ends, so that it geometrically resembles a ring. We formulate and solve the eigenvalue problem in Eqn.~\eqref{eq: eig discrete} as a function of the projection parameter $\theta$, i.e. for $\bK=\bK(\theta)$. The following exponential basis is employed as comparison functions in Eqn.~\eqref{eq: Galerkin}:
\begin{equation}\label{ring basis}
 \phi_n(x) = \exp \left(i \dfrac{2 \pi n}{S a} x \right), \quad n=-N,...,+N,
\end{equation}
As we compute the spectrum of the ring structure, we need to ensure that its natural frequencies represent the spectrum of the bulk, \emph{i.e.} of the underlying infinite domain. To this end, we consider a rational $\theta=p/q$, with $p$ and $q$ being coprime integers. The corresponding beam has $q$ springs in a periodic unit cell. The motion of such infinite, periodic beam can be described in terms of Bloch modes $w_{Bl}$, which satisfy the following conditions
\begin{equation}\label{Bloch}
w_{Bl}(x+qa) = e^{i\mu} w_{Bl} (x), \;\; \mu\in [0,2\pi]. 
\end{equation}
Next, we consider a finite beam with $S$ springs, of length $L=aS$, corresponding to an integer number $R$ of unit cells, i.e., $S=qR$. For periodic boundary conditions applied to the ends of the beam, each of its vibration modes $w_r(x)$ is such that
\begin{equation}\label{ringConstraint}
w_{r}(x+aS) = w_{r}(x) . 
\end{equation}
Equation~\eqref{Bloch} implies
\begin{equation}\label{Bloch1}
w_{Bl}(x+a S) = w_{Bl}(x+a q R) = e^{i R \mu} w_{Bl}(x). 
\end{equation}
Comparison of Eqn.~\eqref{ringConstraint} with Eqn.~\eqref{Bloch1} above shows that the modes of the rings coincide with the Bloch modes, \emph{i.e.}  $w_{Bl}(x)=w_{r}(x)$ for $x\in [0,Sa]$, if $e^{i R \mu} = 1$, which leads to the following condition on the wavenumber:
\begin{equation}\label{muvalues}
\mu_r =\dfrac{ 2 \pi r }{R} , \;\;\; r = 1,2, ... R. 
\end{equation}
Thus, considering a ring of finite length discretizes the wavenumber into values $\mu_r$ separated by intervals $\Delta \mu=2\pi / R$. At these wavenumbers, the vibration modes computed for the ring are a subset of the Bloch modes of the infinite periodic beam, and the eigenfrequencies $\Omega$ of the ring are a discrete subset of the bulk spectrum. The density of this subset increases with the number of units $R$ considered, or equivalently, given the projection parameter $\theta=p/q$, with the number of springs $S=q R$ in the finite ring.

For convenience, we compute the bulk spectra of the beams as a function of $\theta$ by considering a ring with a fixed number of springs $S$. To that end, we need to determine the set of $\theta$ values that corresponds to commensurate rings, so that, by the conditions stated above, the eigenfrequencies of the considered structures represent the spectrum of the bulk. The condition is satisfied for rational values of $\theta=p/q$, if the number of springs satisfies $S=qR$, such that the structure is commensurate with $R$ unit cells, which gives
\begin{align}
\theta &=  p/q = p R/S , \;\;\; p,R \in \mathbb{Z} \nonumber \\ &= s/S, \;\;\; s = 1,2,...S.  \label{theta_commensurate} 
\end{align}
The last equality follows from restricting $\theta\in[0,1]$ and noting that the multiplication of integers modulo $S$ forms a group. Indeed, we can always find a unique integer $s\in\{1,2,...,S\}$ satisfying $s \equiv p R \mod(S) $ for given $p,R\in \mathbb{Z}$.
Thus, we can investigate the spectral properties of infinite beams as a function of the projection parameter $\theta$ by discretizing its range in steps $\Delta \theta =1/S$, which leads to an infinitely dense subset of $[0,1]$ as $S\rightarrow\infty$. This discretization identifies all commensurate, or periodic, rings defined by rational values of $\theta$, whose vibrational spectrum approximates the bulk spectrum evaluated for all $\theta \in [0,1]$.

\begin{figure}[b!]
\centering
\includegraphics[width=0.46\textwidth]{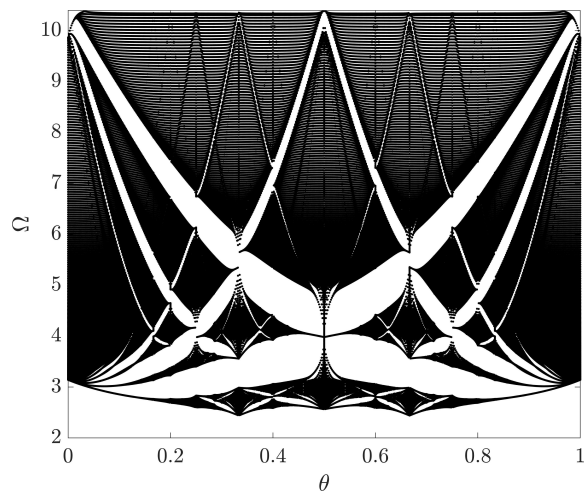}
\caption{Bulk spectrum of quasiperiodic beam lattice with $\gamma_g=10$ showing fractal bandgaps as $\theta$ varies.}\label{beambulk}
\end{figure}

The investigation of the bulk spectrum approximated as described above leads to the results displayed in Fig.~\ref{beambulk}. The calculations are conducted for $S=600$, $\gamma_g=10$ and a set of basis functions corresponding to $N=1000$ in Eqn.~\eqref{ring basis}. The black regions define ranges of frequency populated by the bulk eigenvalues, while the white areas correspond to frequency ranges where no states exist and identify bandgaps. The spectrum has features similar to the Hofstadter butterfly spectrum encountered in quantum mechanics for lattices under a magnetic field~\cite{hofstadter1976energy} and in discrete mechanical QP lattices~\cite{apigo2018topological,martinez2018quasiperiodic}. One can observe the existence of a low frequency bandgap starting at zero, due to the presence of the ground springs, and a number of other gaps associated with Bragg scattering. As $\theta$ increases from $0$, the bulk bands split into a series of smaller bands forming several gaps which presents the fractal structure typical of QP media~\cite{hofstadter1976energy}.

\begin{figure}[h!]
\centering
\subfigure[]{\label{figtheta0}
\includegraphics[width=0.46\textwidth]{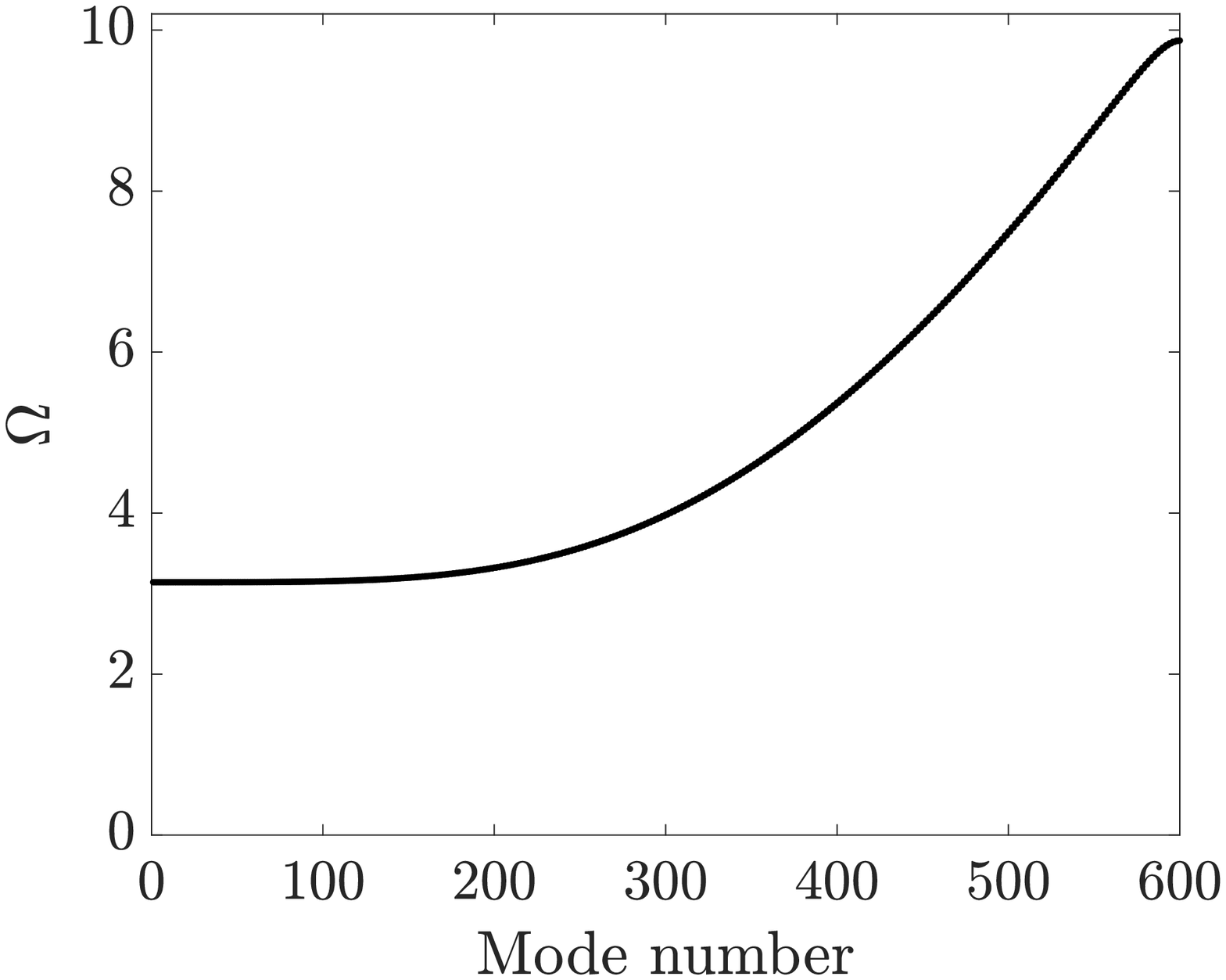}
}
\subfigure[]{\label{figp1q3}
\includegraphics[width=0.46\textwidth]{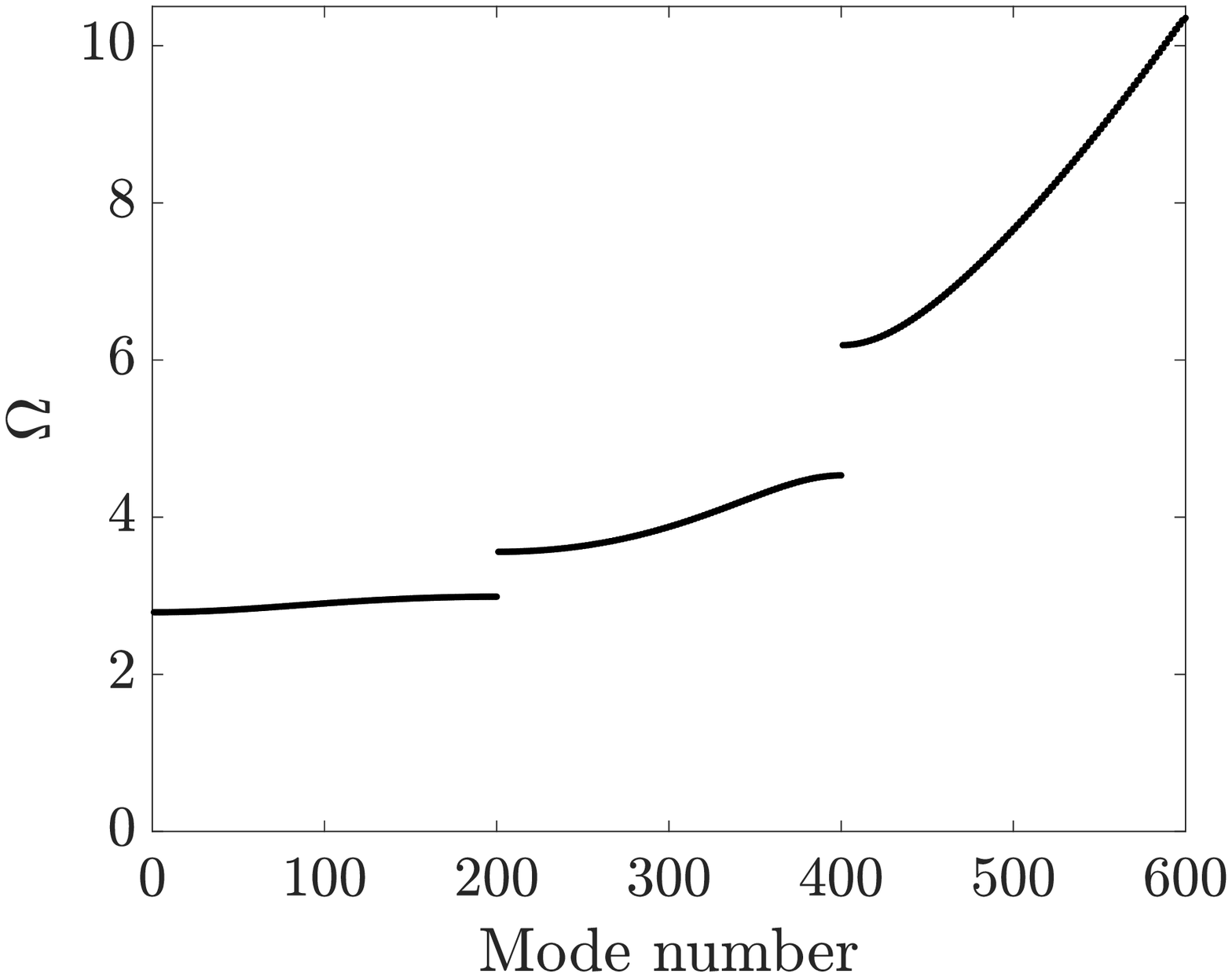}
}
\subfigure[]{\label{figp1q5}
\includegraphics[width=0.46\textwidth]{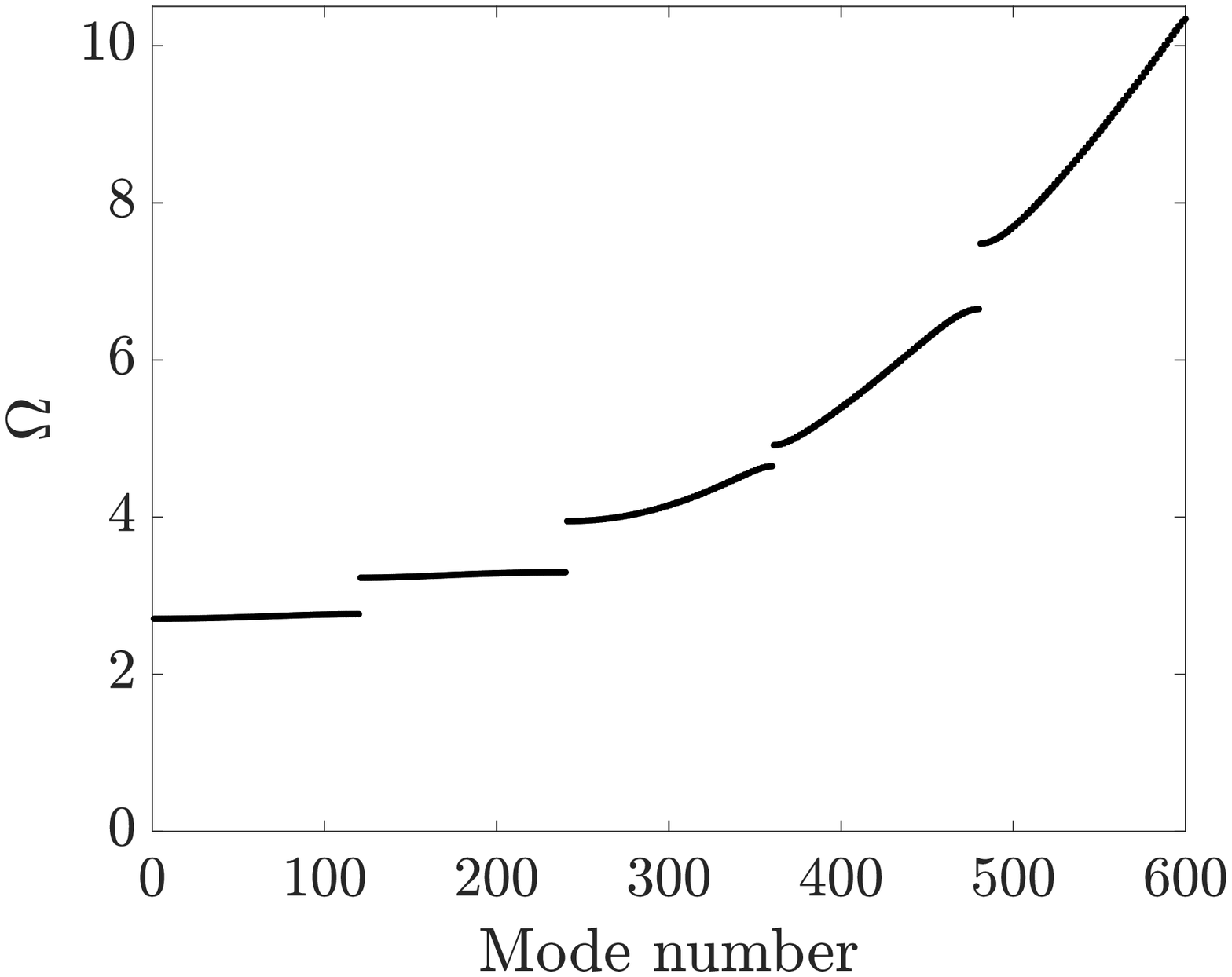}
}
\caption{Frequency spectra for a beam with $S=600$ and $\gamma_g=10$. The first $600$ eigenfrequencies initially in a single band for $\theta=0$ (a) split into three bands for $\theta=1/3$ (b) and into five bands for $\theta=1/5$ (c). 
} 
\end{figure}

We focus on the spectrum formed by the first $S=600$ modes to illustrate the topological properties of the bandgaps and the resulting vibration modes localized at the boundaries of finite beams. Such spectrum captures the splitting of the first band at $\theta=0$ into several other bands as $\theta$ varies. For example, Fig.~\ref{figtheta0} shows the first $S=600$ modes for $\theta=0$, where all the modes lie on a single band, while for $\theta=1/3$ these modes split into three bands (Fig.~\ref{figp1q3}), and for $\theta=1/5$ they split into five bands (Fig.~\ref{figp1q5}). In general, for a rational $\theta=p/q$, where $p$ and $q$ are coprime, we expect the first $S$ modes to be distributed among $q$ bands with at most $q-1$ bandgaps between them. The spectrum of Fig.~\ref{beambulk} captures this behavior for all the $\theta$ values that are commensurate with $S=600$, as discussed previously, and is an approximation of the continuous spectrum formed by all real $\theta \in [0,1]$. 

\begin{figure}
	\centering
	\subfigure[]{\label{IDSfig_full}
\includegraphics[width=0.46\textwidth]{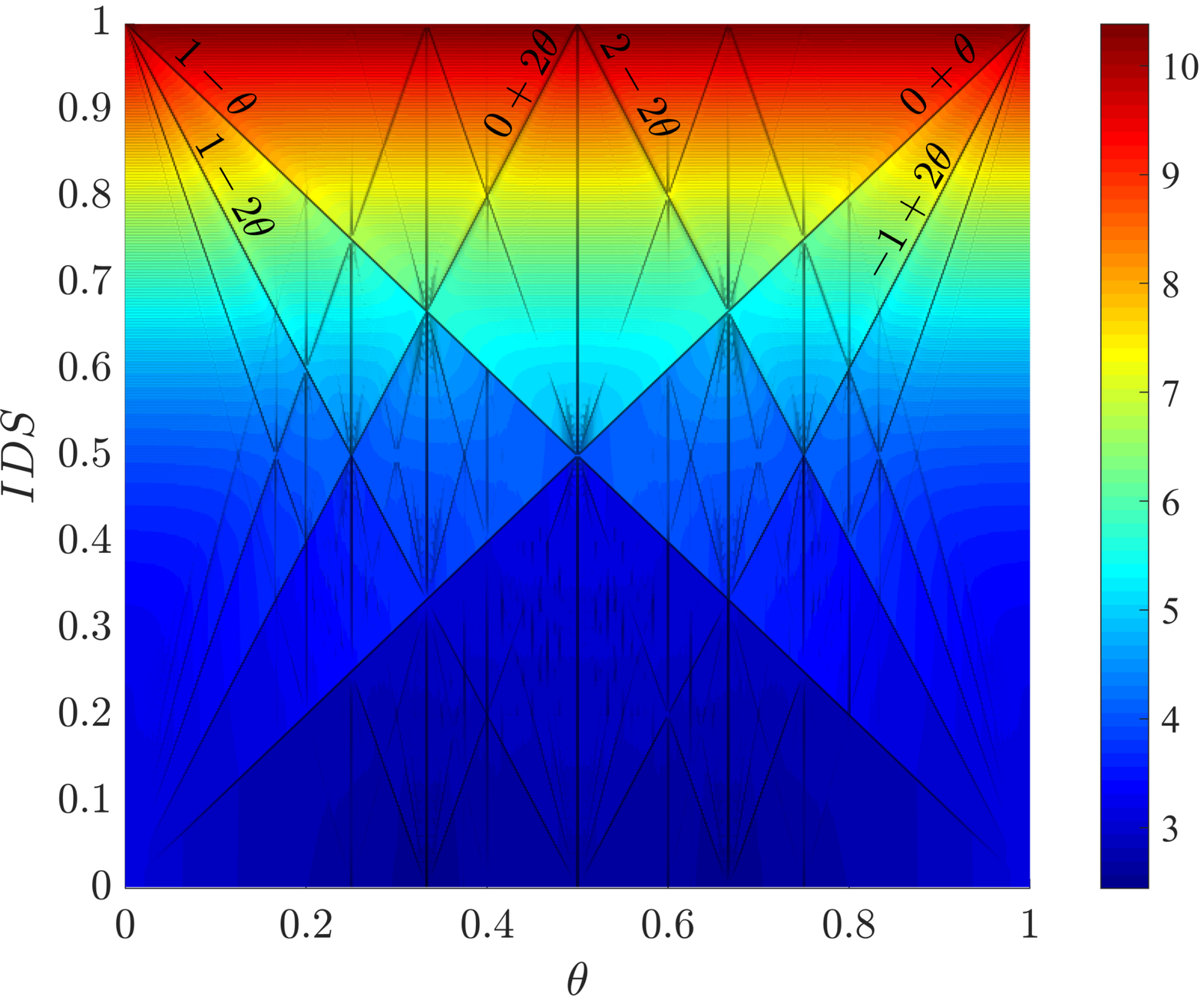}
}	
\subfigure[]{\label{IDSfig_p1q5}
		\includegraphics[width=0.45\textwidth]{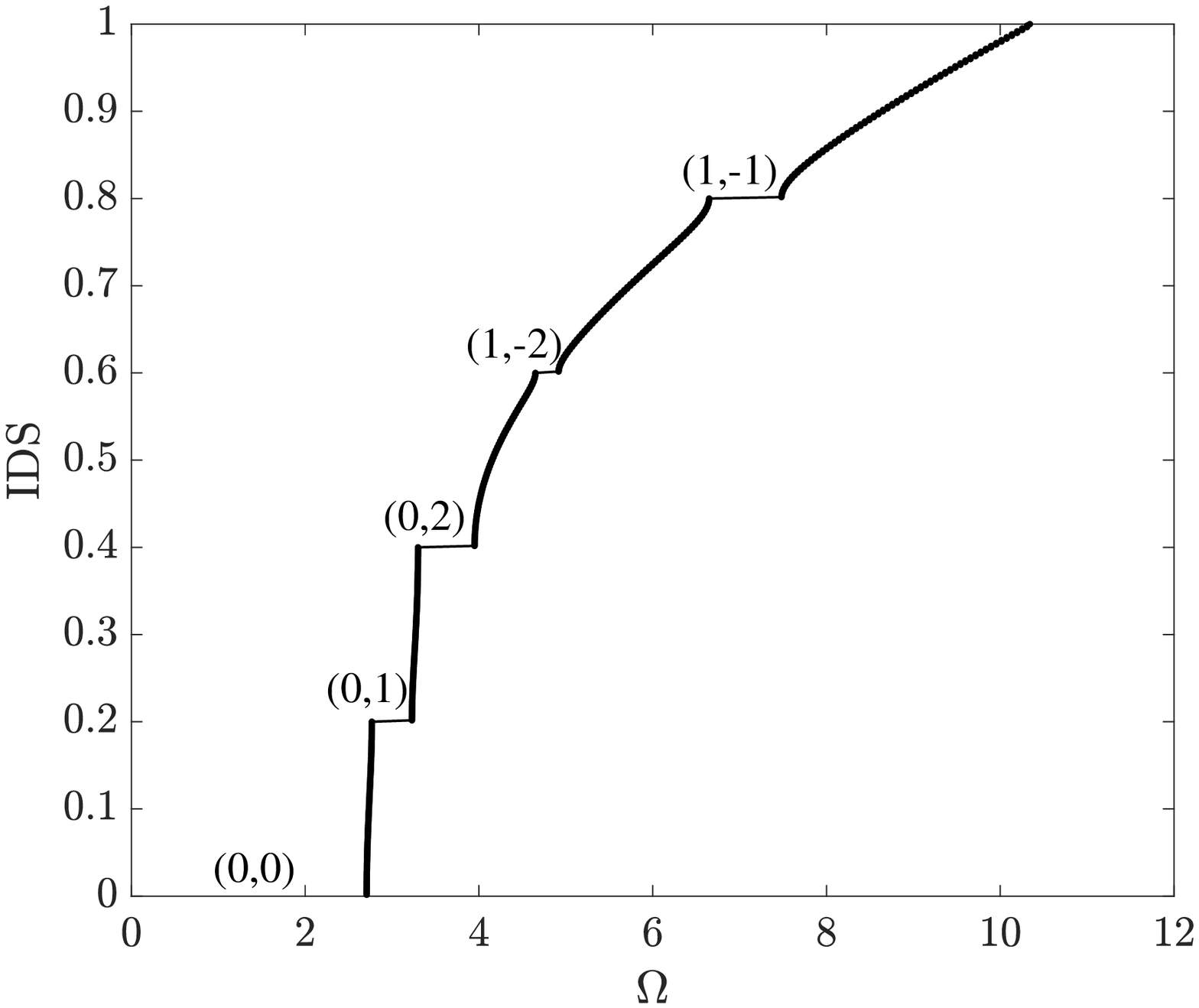}
	}
	\caption{(a) Integrated density of states (IDS) with $\theta$ exhibiting sharp jumps at the bandgaps. The colormap indicates frequency $\Omega$. (b) IDS at $\theta=1/5$ along with the topological invariants (n,m) at each bandgap.} 
	\label{FigSpectra}
\end{figure}

Splitting of the bulk bands corresponds to a change in density of states, which leads to the topological classification of the bandgaps~\cite{bellissard1986k,prodan2018k}. The integrated density of states (IDS) at frequency $\Omega$ is defined as:
\begin{equation}\label{IDS_omeg Defn.}
\text{IDS} (\Omega) = \lim_{S\to \infty} \dfrac{\sum_n [\omega_n\leq \Omega ]}{S}, 
\end{equation}
where $[\cdot]$ denotes the Iverson Brackets, which provide a value of $1$ whenever the argument is true. 

Figure~\ref{IDSfig_full} displays the IDS as a function of $\theta$, corresponding to the bulk spectrum data of Fig.~\ref{beambulk}. Recall that we restrict attention to the spectrum formed by the first $S=600$ modes, and thus the IDS lies in the range $\{ 0,1 \}$. The colormap represents frequency $\Omega$, and the rendering highlights the sharp changes that indicate frequency jumps of the IDS inside the bandgaps. These jumps occur since all the frequencies inside a bandgap have the same IDS. The fact that the IDS inside each gap is characterized by a straight line is a consequence of the pattern defined by the projection from a circle~\cite{apigo2018topological,prodan2016bulk}. For such a pattern, the IDS of bandgap $g$ can be expressed as
\begin{equation}\label{IDS_invariant}
\text{IDS}(g) = n + m \theta,
\end{equation} 
where $\{n,m \}\in \mathbb{Z}$ are invariant labels of the bandgap~\cite{bellissard1986k,prodan2018k}. A few lines linked to the most prominent gaps have their IDS equation displayed in Fig.~\ref{IDSfig_full}. In particular, the slope $m$ gives the number of topological edge modes spanning the corresponding gap in the spectrum of a finite structure. For example, for $p/q=1/5$ (Fig.~\ref{figp1q5}) the four bandgaps are characterized respectively by $IDS=\theta$, $IDS=2\theta$, $IDS=1-2\theta$ and $IDS=1-\theta$ (Fig.~\ref{IDSfig_p1q5}).

\subsection{The spectrum of finite QP beams}\label{Sec: finiteQP}
We now examine the spectrum of finite QP beams and its relation to the bulk spectrum. We show that the finite size and the presence of boundaries produces additional modes that are localized at one boundary, and whose frequencies span the bulk bandgaps as the projection parameter $\theta$ varies. In addition, we illustrate how these modes migrate and localize in the interior of the domain as their frequencies merge with the bulk bands.

We consider a beam of length $L=a S_f$ comprising of $S_f-1$ springs and simply supported at both ends, \emph{i.e.} subjected to the following boundary conditions $w(x=0,L) = 0,  w_{xx}(x=0,L) = 0$. Quasiperiodic patterns are generated by considering projections from circles centered at $x_s \in \{a,2a,...,(S_f-1)a\}$. The finite beam eigenfrequencies computations are conducted through the Galerkin's approximation, with the following set of comparison functions:
\begin{equation}\label{sineBasis}
\phi_n(x) = \sin \left( \dfrac{\pi n x}{a S_f }\right), \quad n=1,..,N
\end{equation}

\begin{figure}
	\centering
	\begin{minipage}{0.53\textwidth}
	\subfigure[]{
		\includegraphics[width=1\textwidth]{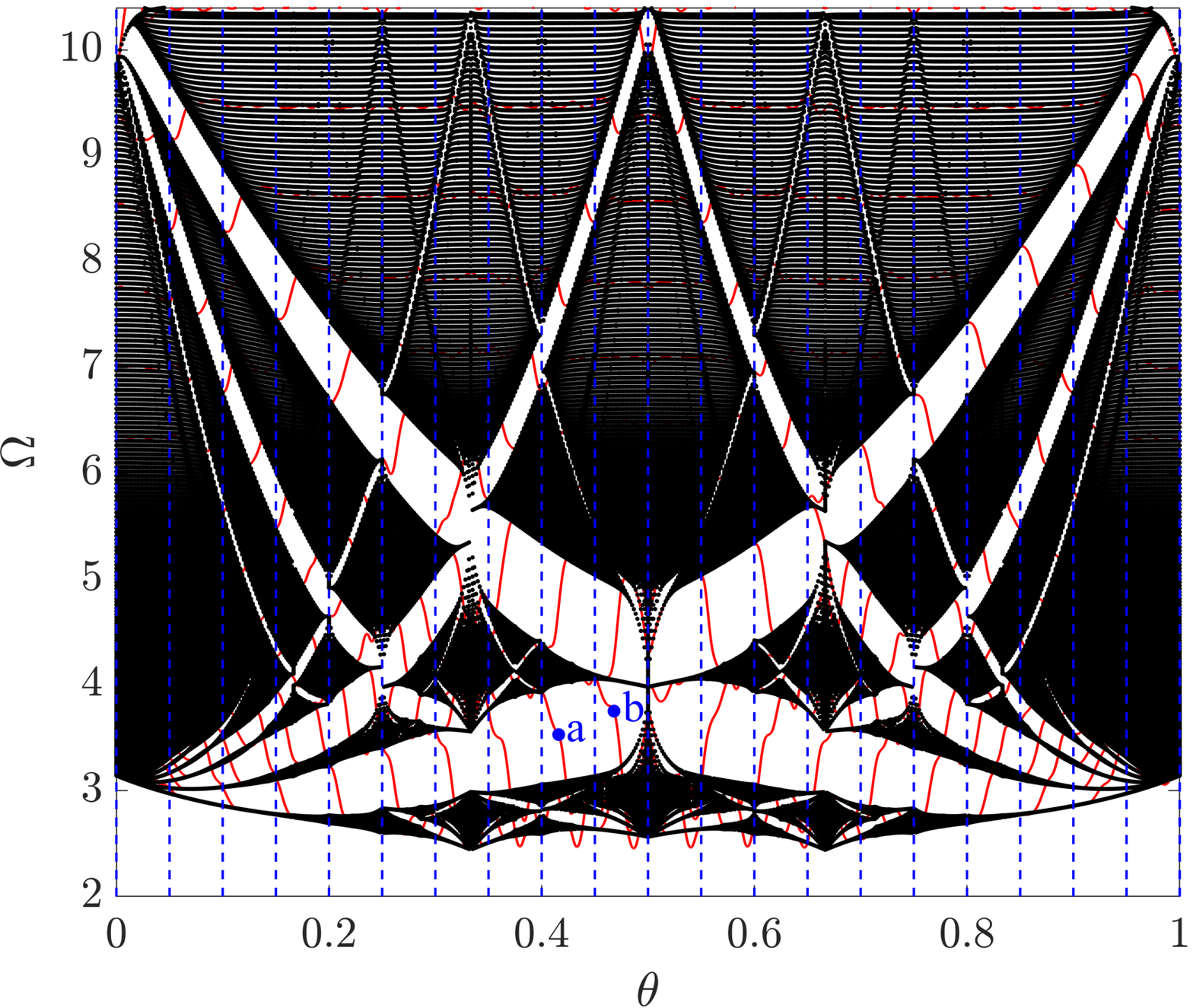}\label{beamFiniteSpectrumFig}
	}
	\end{minipage}\hspace{0.3cm}
	\begin{minipage}{0.43\textwidth}
	\subfigure[]{
		\includegraphics[width=1\textwidth]{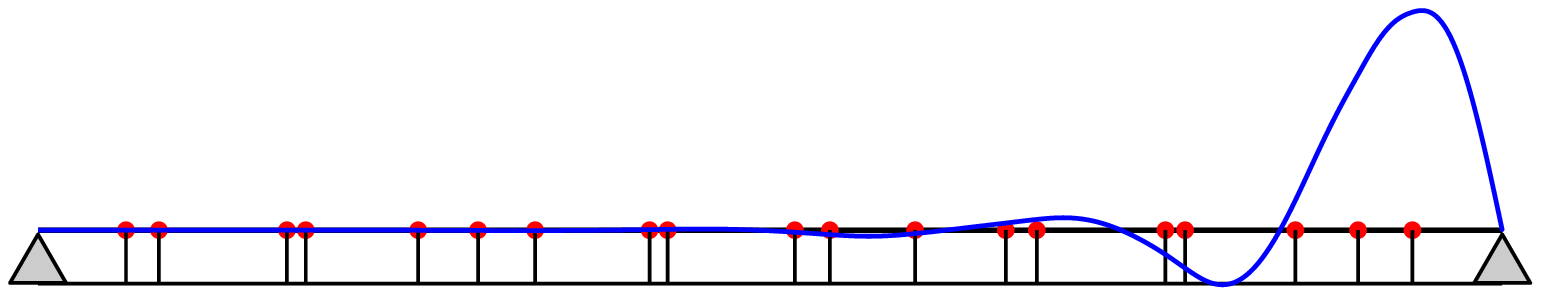}
		\label{modeA}
	}\\
	\subfigure[]{
		\includegraphics[width=1\textwidth]{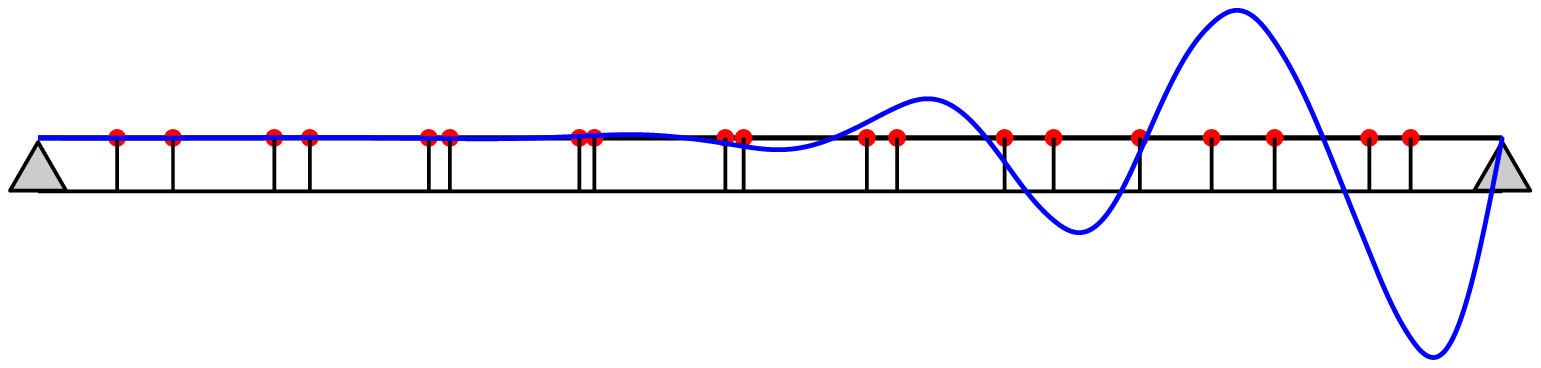}\label{modeB}
	}
	\end{minipage}
	\caption{(a) Spectrum of finite beam with $S_f=20$ (red) superimposed by the bulk spectrum (black) as a function of $\theta$. Red curves spanning the bulk gaps are topological edge modes localized at the right boundary of the finite beam. (b,c) Mode shapes corresponding to the eigenfrequencies marked by blue dots with labels a and b, respectively.} \label{finitespectrumfig}
\end{figure}

The eigenfrequencies for a finite beam with $S_f=20$ are computed for varying $\theta$ and superimposed (in red) to the bulk spectrum (in black) in Fig.~\ref{beamFiniteSpectrumFig}. The computations are conducted using $N=700$ basis functions of the kind above. Results show that the spectrum of the finite beam is generally superimposed to the bulk, while additional modes spanning the gaps are generated by the finite length of the domain and the presence of the considered boundaries. These additional modes are localized at the right boundary, as shown in the examples displayed in Figs.~\ref{finitespectrumfig}(b,c). Insight into their nature, their relation with the IDS and its topological invariants is given through the observations presented below. Similar considerations are made for the simple case of a discrete spring mass chain described in the Appendix, which is introduced as an additional aid to the understanding of the discussion below. 

{
The existence of edge states spanning the gaps is related to changes in the density of states with $\theta$, which for each bandgap is quantified by the slope $m$ of the corresponding IDS line. To illustrate this connection, we first note that the finite beam spectra overlaps with that of the bulk for values of $\theta$ such that $2S_f \theta = p, \hspace{2mm}  p \in \mathbb{Z}$. At such values, the mode shapes of the finite beam of length $S_fa$ can be mirrored at $x=L$ to also satisfy the boundary value problem for a ring structure of size $2S_fa$, with the same eigenfrequency. Hence, since these eigenfrequencies are modes of a commensurate ring, they lie within the bulk. This occurs at
\begin{equation}\label{theta_p}
\theta_p=\frac{p}{2S_f}, \quad p\in \mathbb{Z}\
\end{equation}
Since for these values of $\theta$ the eigenvalues of the finite system lie within the bulk spectra, the condition also holds for the commensurate values of $\theta$ given by Eqn.~\eqref{theta_commensurate}, \textit{i.e.} $\theta_s=s/S_f, \hspace{2mm} s=1,2,... S_f$, since they are a subset of the values given by Eqn.~\eqref{theta_p}. Dashed vertical blue lines mark these commensurate values of $\theta$ in Fig.~\ref{beamFiniteSpectrumFig}.}

Next, we evaluate the number of boundary modes that span a bandgap and the relation to the respective gap labels. Based on the discussion above, we investigate the behavior of a gap in the interval between two subsequent commensurate values of $\theta$, namely $\theta_i, \theta_{i+1}$ corresponding to $s=i, i+1$. As noted above, at the commensurate $\theta_s$ values, the frequencies of the finite system all lie within the bulk spectra, thus the bulk IDS quantifies the number of eigenstates of the finite system below the bandgap of interest. Therefore, using Eqn.\eqref{IDS_invariant} we can quantify the finite change $\Delta IDS$ occurring for $\Delta \theta =\theta_{i+1}-\theta_i$ to be:
\begin{equation}\label{nmodes}
\Delta IDS  = m (\theta_{i+1}-\theta_i) = m\left( \frac{i+1}{S_f}-\frac{i}{S_f} \right) = \frac{m}{S_f}
\end{equation} 
where $m$ is the bandgap topological invariant. From the definition of the IDS, the change in the number of modes $N_g$ below the gap between $\theta_i$ and $\theta_{i+1}$ is 
\begin{equation}\label{nmodes}
N_g  = S_f \Delta IDS = m.
\end{equation} 

A key observation is that each of the eigenfrequencies of the finite beam vary continuously with $\theta$ since the stiffness matrix $\bK$ is a continuous function of $\theta$. Therefore, the only way the number of modes below a gap can change between two commensurate values of $\theta$ is through a mode migration from a bulk band to its adjacent one. In particular, positive $m$ values signal modes migrating from the band above to the band below the gap, while negative $m$ values signal an opposite behavior. These statements can be verified in Fig.~\ref{beamFiniteSpectrumFig}. The larger gaps are those for which $|m|=1$, and one can note that between any two commensurate values of $\theta$ (dashed blue lines) a single mode is spanning these gaps. For smaller gaps, such as those with $|m|=2$ or $|m|=3$, we also observe two or three spanning modes, respectively. Also, $m>0$ signals a mode migration from the top to the bottom band of the gap with increasing $\theta$, while an opposite behavior is observed for $m<0$. The numerical observations are therefore in full agreement with the predictions based on the gap invariants defined from the IDS. 

We now examine the shapes of the modes that are transversing the bandgaps. For example, the modes marked by blue dots in Fig.~\ref{beamFiniteSpectrumFig} are displayed in Fig.~\ref{modeA} and Fig.~\ref{modeB}. Since their eigenfrequencies lie inside a bandgap, these modes cannot be globally spanning bulk modes, and therefore are localized at a boundary. Moreover, all of the modes spanning the gaps are localized at the right boundary of the structure. This is a consequence of the way a finite structure is constructed from the pattern defined previously, that is, how an infinite pattern is terminated to create a finite pattern. Here, a finite structure is constructed by starting at $x=0$ (left boundary) and adding springs to the right boundary. Therefore, for a given $\theta$, a finite beam can be seen as a cut of a larger commensurate structure that could be created by continuing to add springs to the right boundary until $S_f \theta$ is an integer. For such commensurate structure, all the eigenfrequencies define bulk modes. By terminating it before $S_f \theta$ is an integer, a localized mode appears at the right boundary, where the cut is made. In the next section, we show how localized modes at the left boundary can be obtained by introducing a phase parameter in the patterning of the ground springs.

We emphasize that the key properties that result in localized edge states is the incommensurate nature of the lattice, and the properties of the projection rule defining the location of the ground springs. The gap labels guarantee the presence of a number of modes spanning the gaps between two commensurate $\theta$ values, but do not provide any information on the shape of the branch transversing the gap. In fact, one cannot guarantee the presence of a localized mode for any $\theta$ between two commensurate values, since the mode that span the gap can merge with the bulk before or after the considered value. Therefore, these edge modes are better understood by considering the spectrum formed by the whole family of beams with varying $\theta$, instead of single instances of $\theta$.

\begin{figure}[h!]
	\centering
	\subfigure[]{\label{butterflyzoom}
\includegraphics[width=0.475\textwidth]{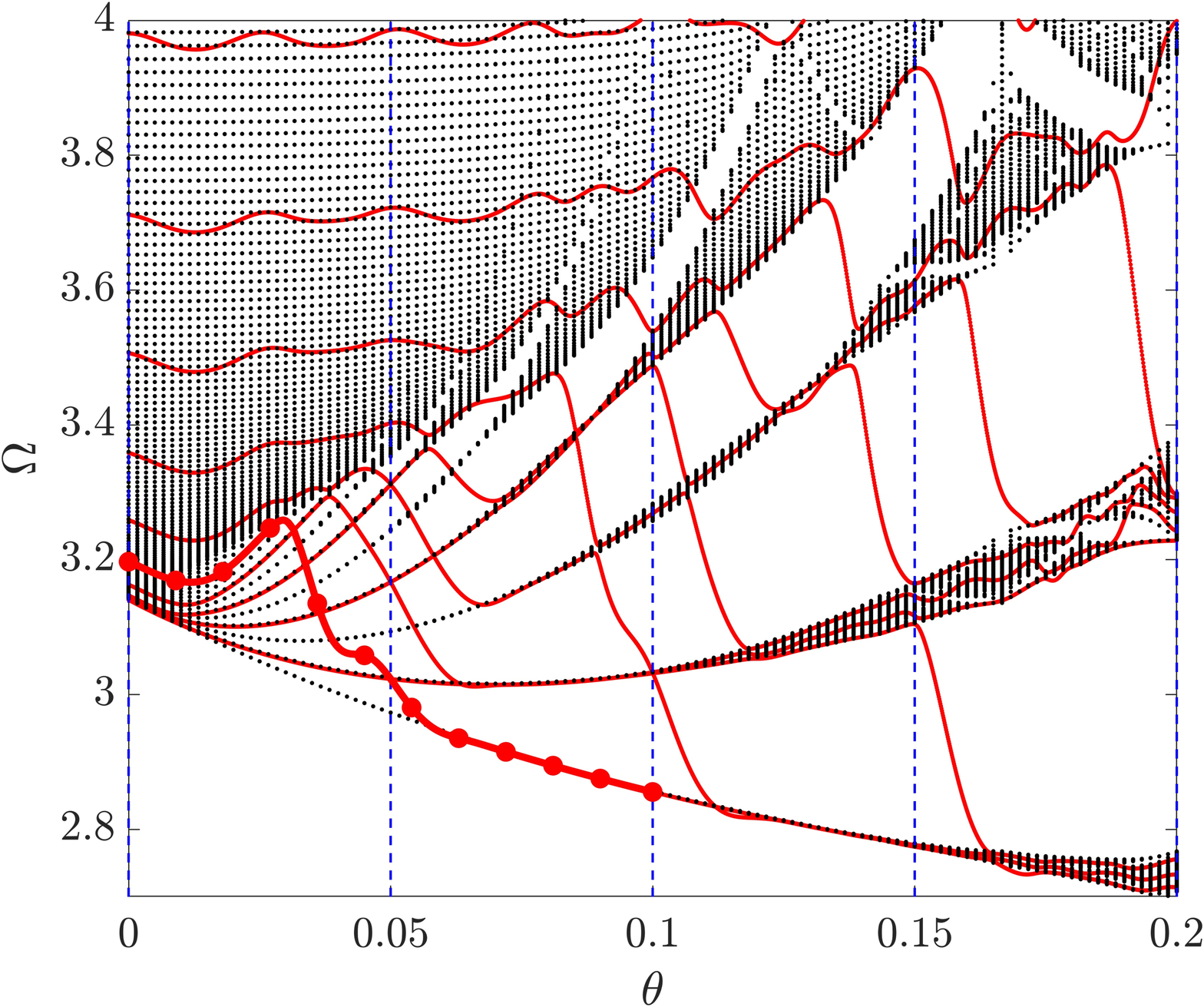}
}	
\subfigure[]{\label{modetransition1}
		\includegraphics[width=0.475\textwidth]{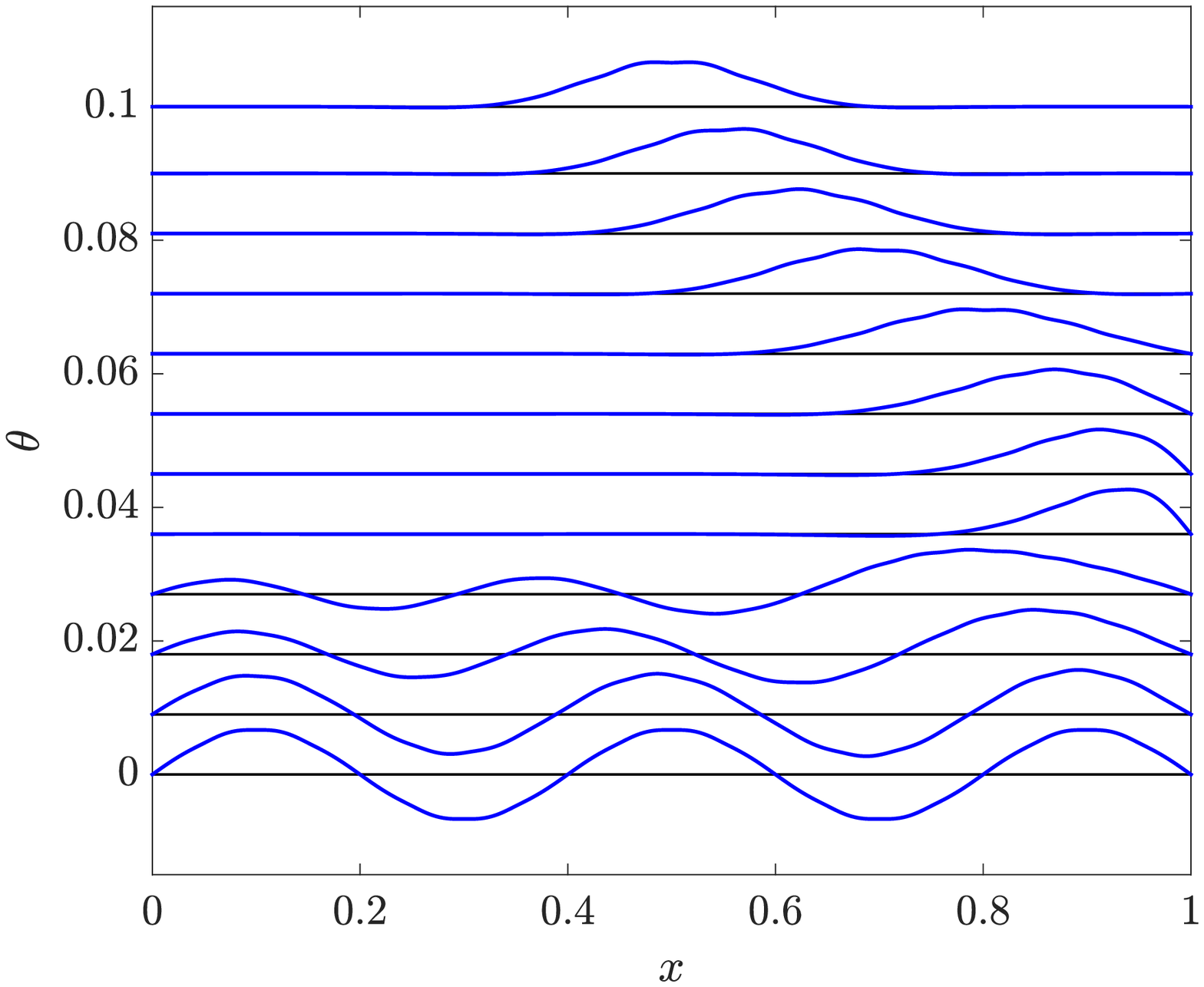}
	}
	\subfigure[]{\label{butterflyzoom2}
\includegraphics[width=0.475\textwidth]{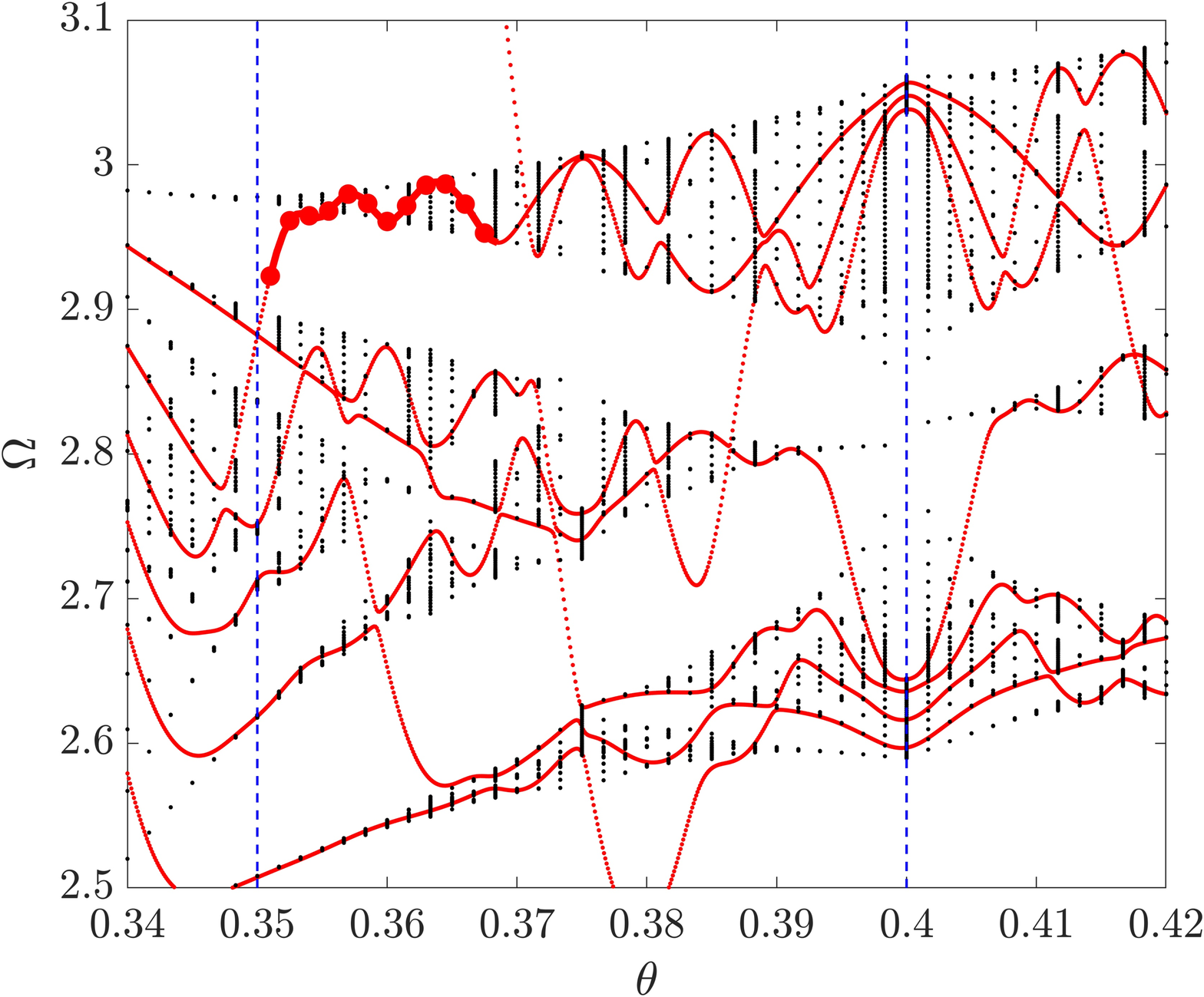}
}	
\subfigure[]{\label{modetransition2}
		\includegraphics[width=0.475\textwidth]{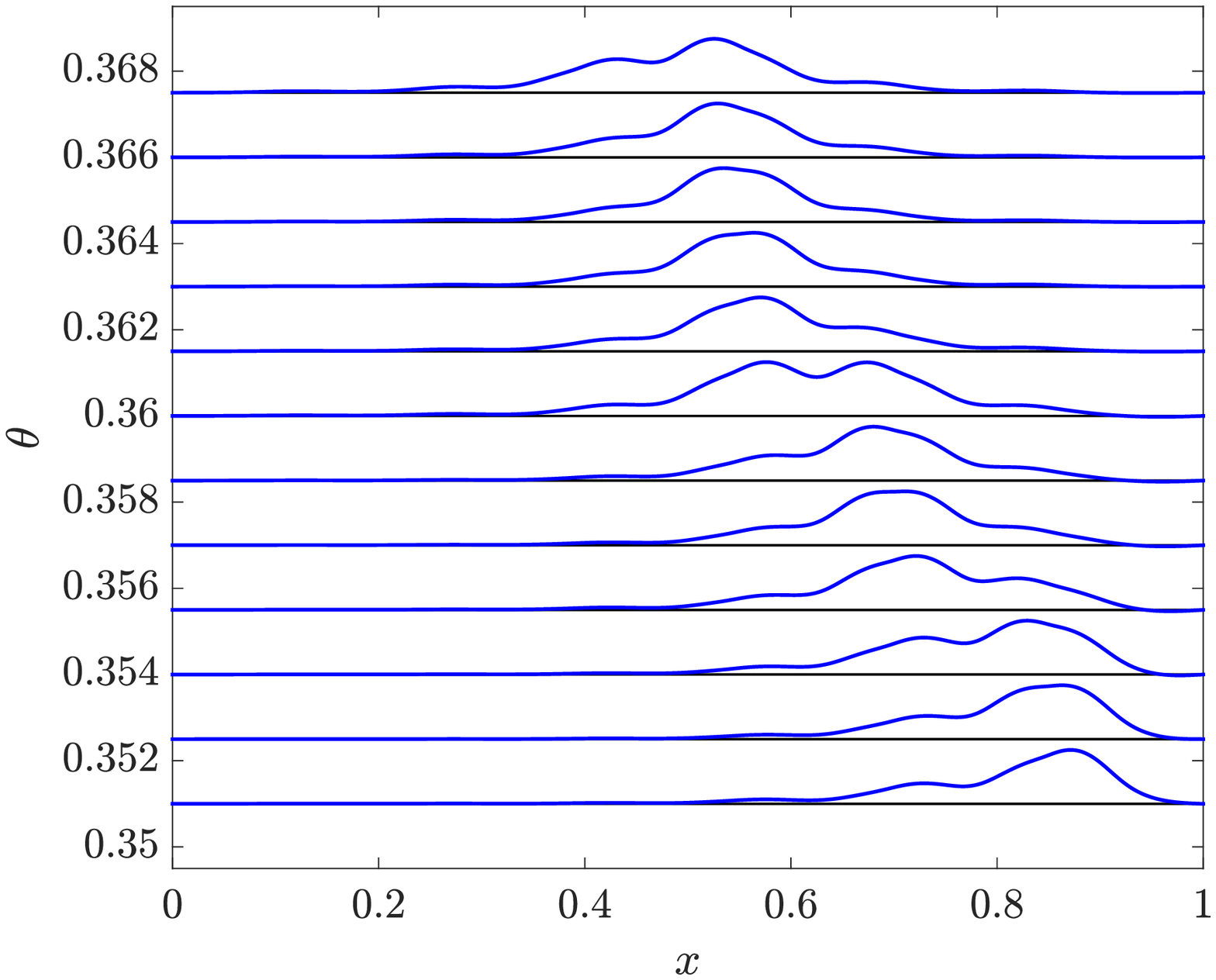}
	}
	\caption{Localization transitions for QP beam with $S_f=20$, $\gamma_g=10$. (a,c) Different zooms of superimposed bulk (black) and finite system spectrum (red) where branches of selected vibration modes are tracked (thick red lines). (b,d) Normalized vibration mode shapes (blue lines) of branches tracked respectively in (a,c) as a function of $\theta$. The set of modes displayed in (b,d) correspond to the red dots marked in the branches of (a,c).} 
	\label{FigInterior}
\end{figure}

In addition to edge states, the family of QP beams supports vibration modes localized in the interior of the domain, whose eigenfrequencies lie inside the regions defined by the bulk bands. Although not classified as topological since they are not spanning non-trivial gaps (characterized by non-zero $m$ labels), these modes arise from the merging of a topological mode branch with the bulk. As an illustrative example, we consider the vibration mode highlighted by the thick red line in the zoomed spectrum of Fig.~\ref{butterflyzoom}, and track how its mode shape evolves with $\theta$ (Fig.~\ref{modetransition1}). The vibration mode starts as a bulk mode for $\theta=0$, transforms into a topological edge mode as its branch tranverses the gaps, and then migrates to the interior of the structure as its branch merges with the bulk. The transition from localization at the boundary to localization in the interior as the mode branch merges with the bulk is a typical behavior observed for many branches of the spectrum. Another example for a different $\theta$ interval is illustrated in Figs.~\ref{FigInterior}(c,d).

\subsection{Mode transitions driven by phase modulations}
\begin{figure}[b!]
	\centering
	\subfigure[]{\label{distfunction}
\includegraphics[width=0.5\textwidth]{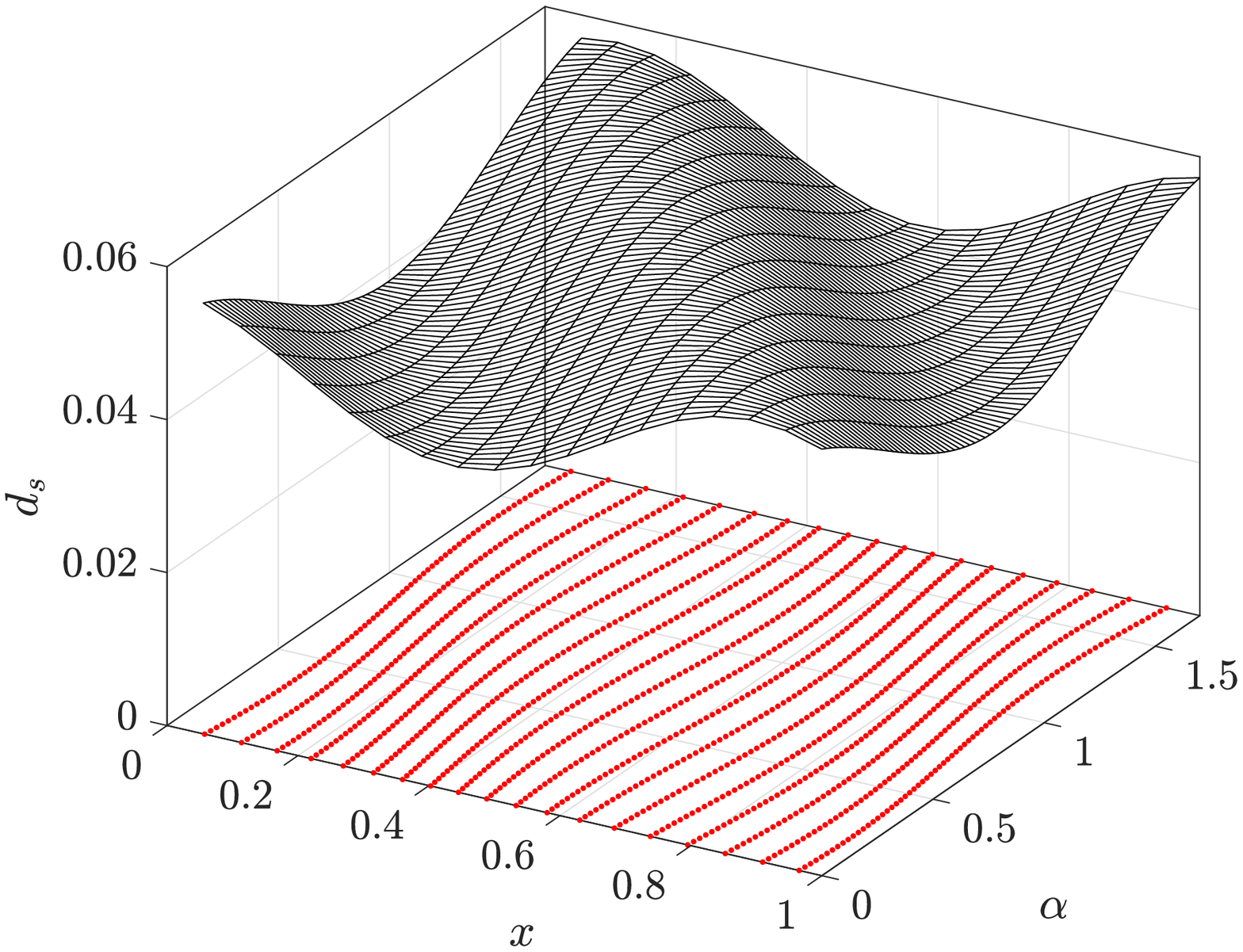}
}
	\subfigure[]{\label{spectrumphase}
\includegraphics[width=0.45\textwidth]{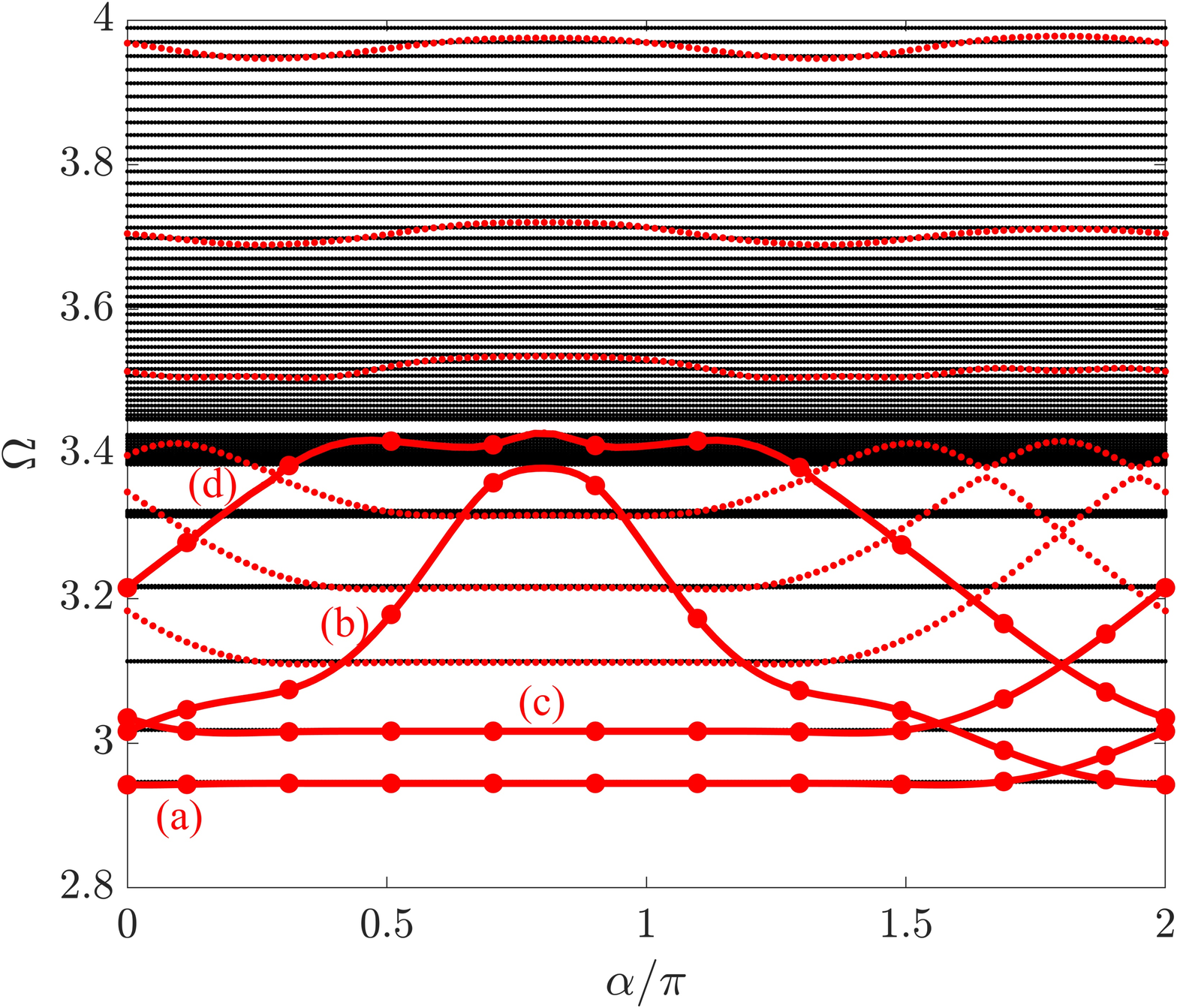}
}
	\caption{Phase modulation for lattice with $\theta=0.06$, $S_f=20$ and $\gamma_g=10$. (a) Location of ground springs $x_s$ (red) and distance between consecutive ground springs $d_s$ (black) as a function of phase $\alpha$. (b) Bulk (black) and finite beam spectra (red) as function of $\alpha$, where selected mode branches are tracked and marked by thick red lines.} 
\end{figure}

The localized vibration modes are here further investigated by introducing a phase parameter $\alpha$ in the QP pattern defining the location of the ground springs. Accordingly, the location of ground spring $s$ is now defined as
\begin{equation}
x_s=sa+R\sin(2\pi s\theta + \alpha).
\end{equation}
It is also useful to define the distance between two consecutive ground springs as $d_s=x_{s+1}-x_s$, which can be expressed as
\begin{equation}
d_s=2R\sin\left(\frac{\theta}{2} \right)\cos\left( 2\pi s\theta + \frac{\pi \theta}{2} + \alpha \right).
\end{equation}
Therefore, the distance between ground springs is modulated with the same spatial frequency $\theta$ and with amplitude $2R\sin(\theta/2)$. For a given $\theta$, varying the phase $\alpha \in [0, 2\pi]$ will result in cyclic modulations of both the ground springs' locations and their distance. These modulations are illustrated for $\theta=0.06$ and $S_f=20$ in Fig.~\ref{distfunction}, where the locations of the ground springs are displayed in red as a function of $\alpha$, and the distance between consecutive ground springs is shown as a black mesh. It is interesting to note that the first vibration mode for $\theta=0.06$ (Fig.~\ref{modetransition1}) is localized at the right boundary, corresponding to a region of low density of springs and a peak of the function $d_s(\alpha=0)$. One may expect that the smooth variation of the ground springs' locations with $\alpha$, and consequently of their distance, will produce a translation of such mode across the domain following the variation of $d_s$. 

\begin{figure}[hb!]
	\centering
\subfigure[]{\label{modephase1}
		\includegraphics[width=0.475\textwidth]{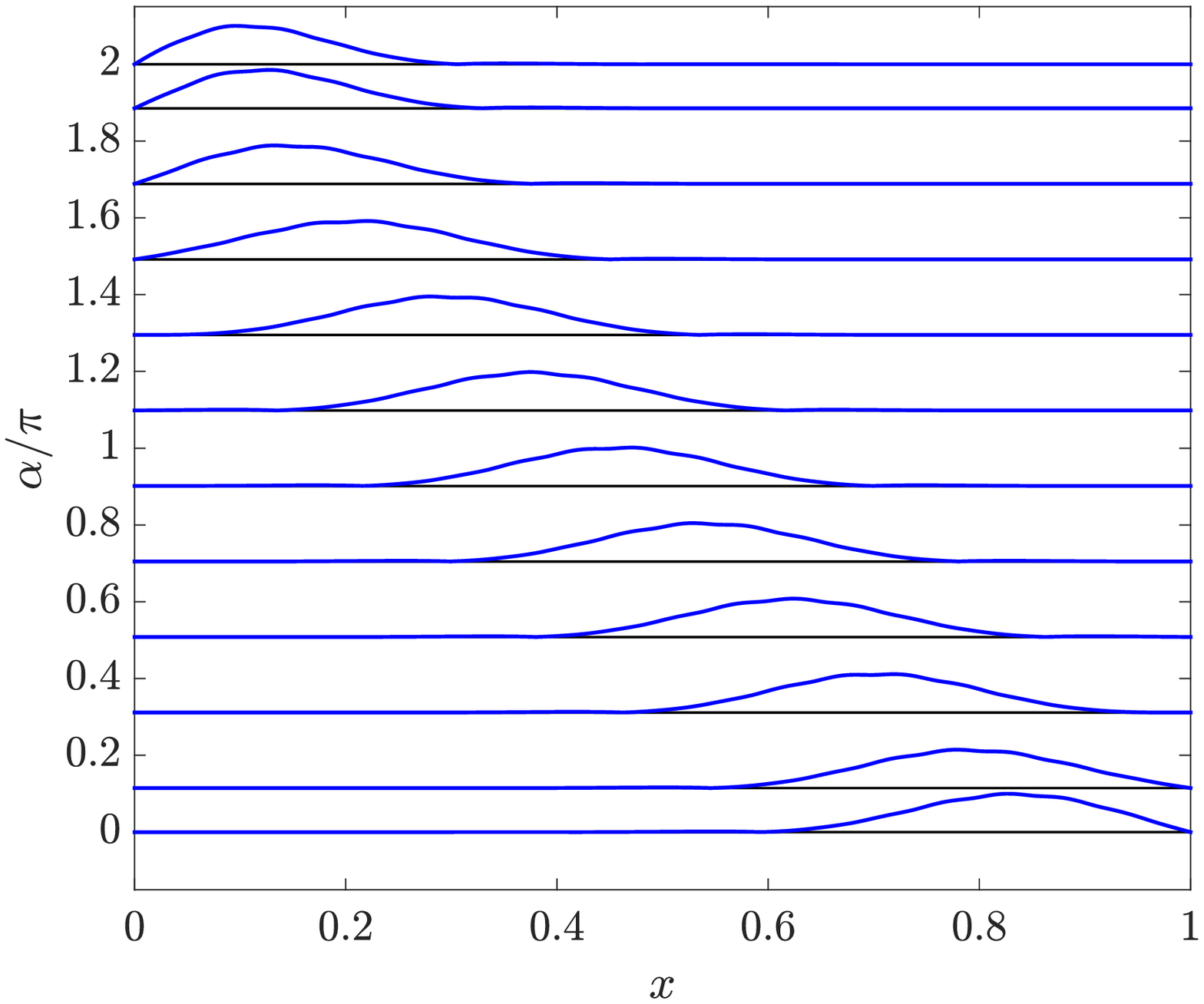}
	}
	\subfigure[]{\label{modephase2}
		\includegraphics[width=0.475\textwidth]{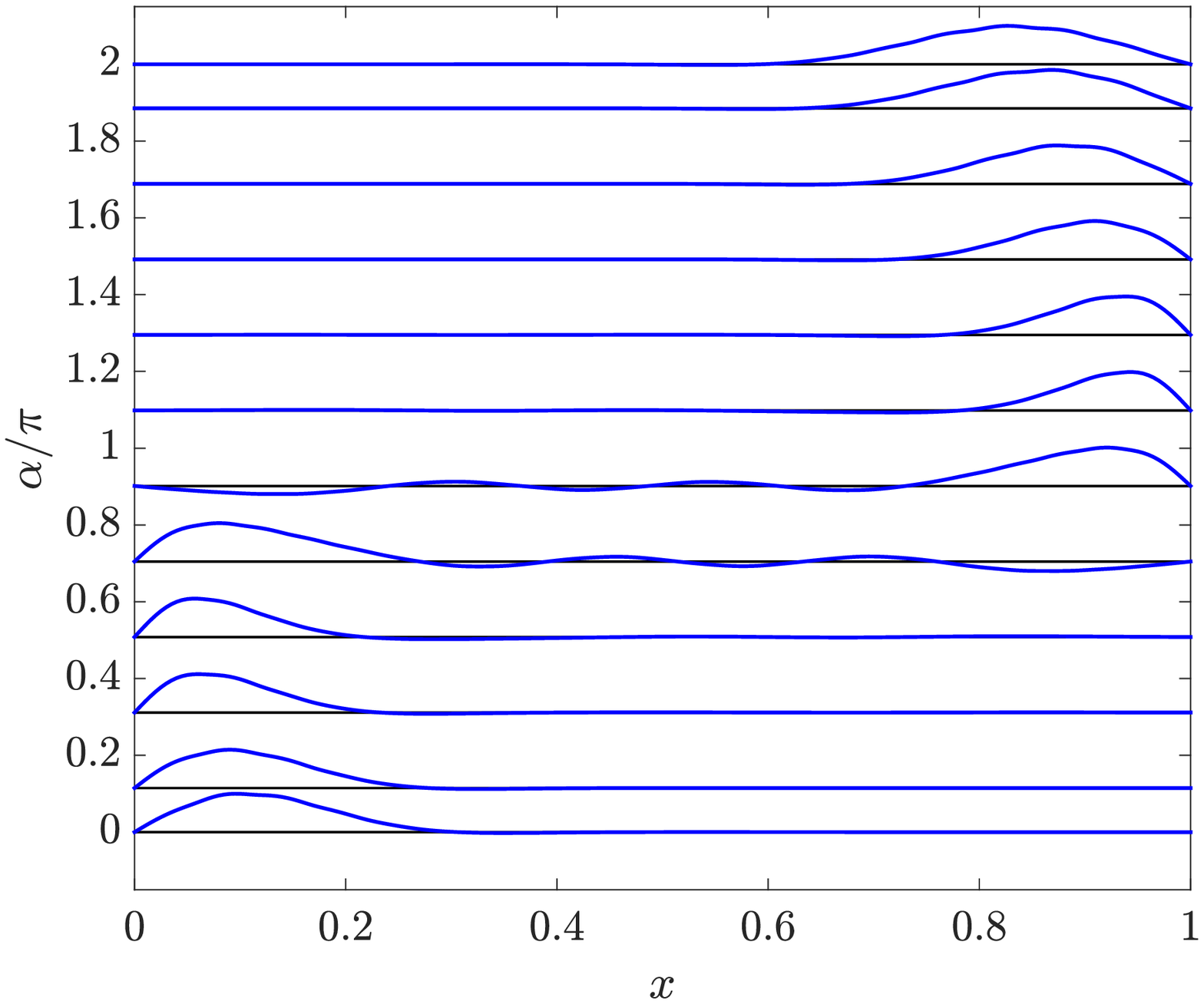}
	}
	\subfigure[]{\label{modephase3}
		\includegraphics[width=0.475\textwidth]{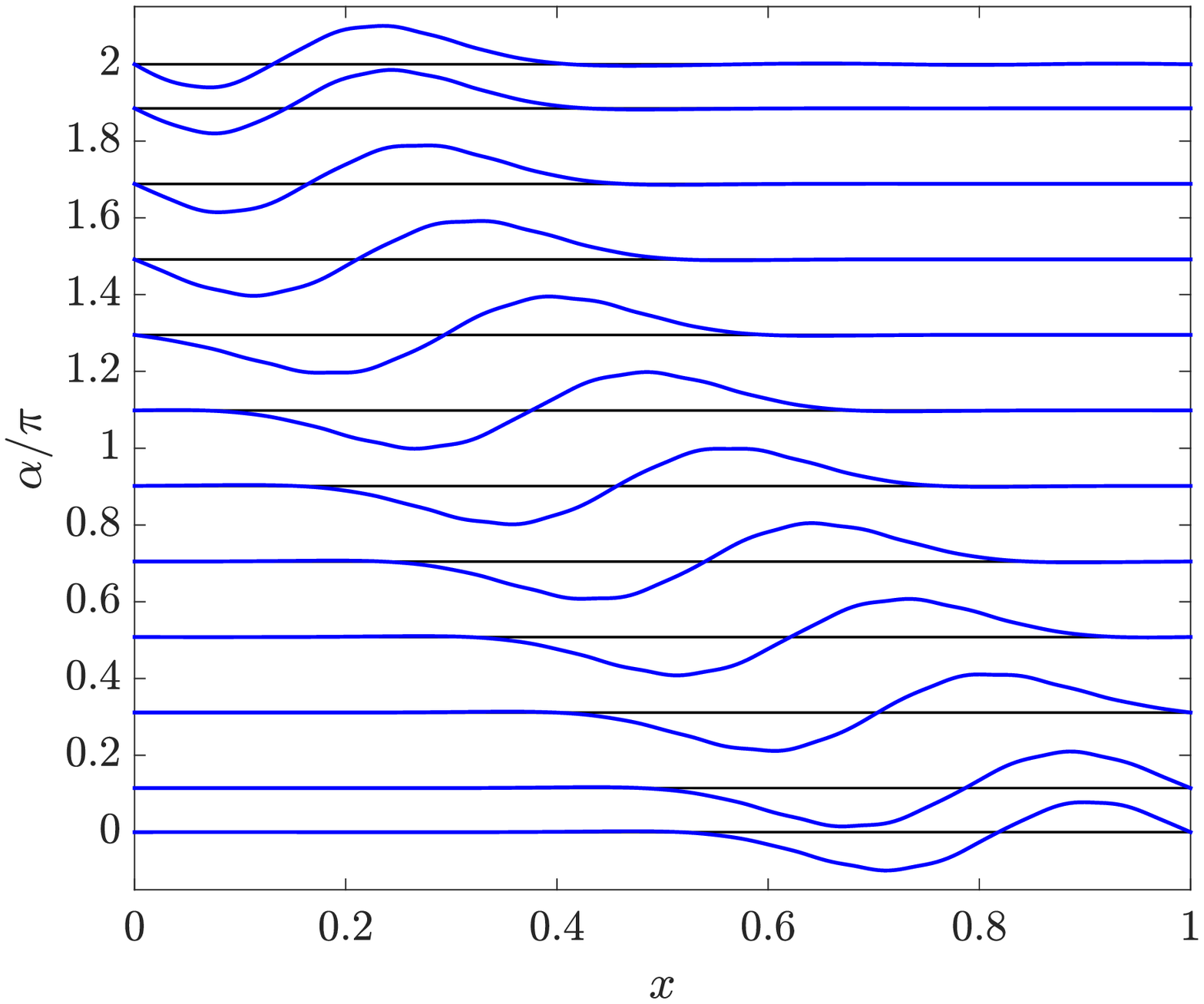}
	}
	\subfigure[]{\label{modephase4}
		\includegraphics[width=0.475\textwidth]{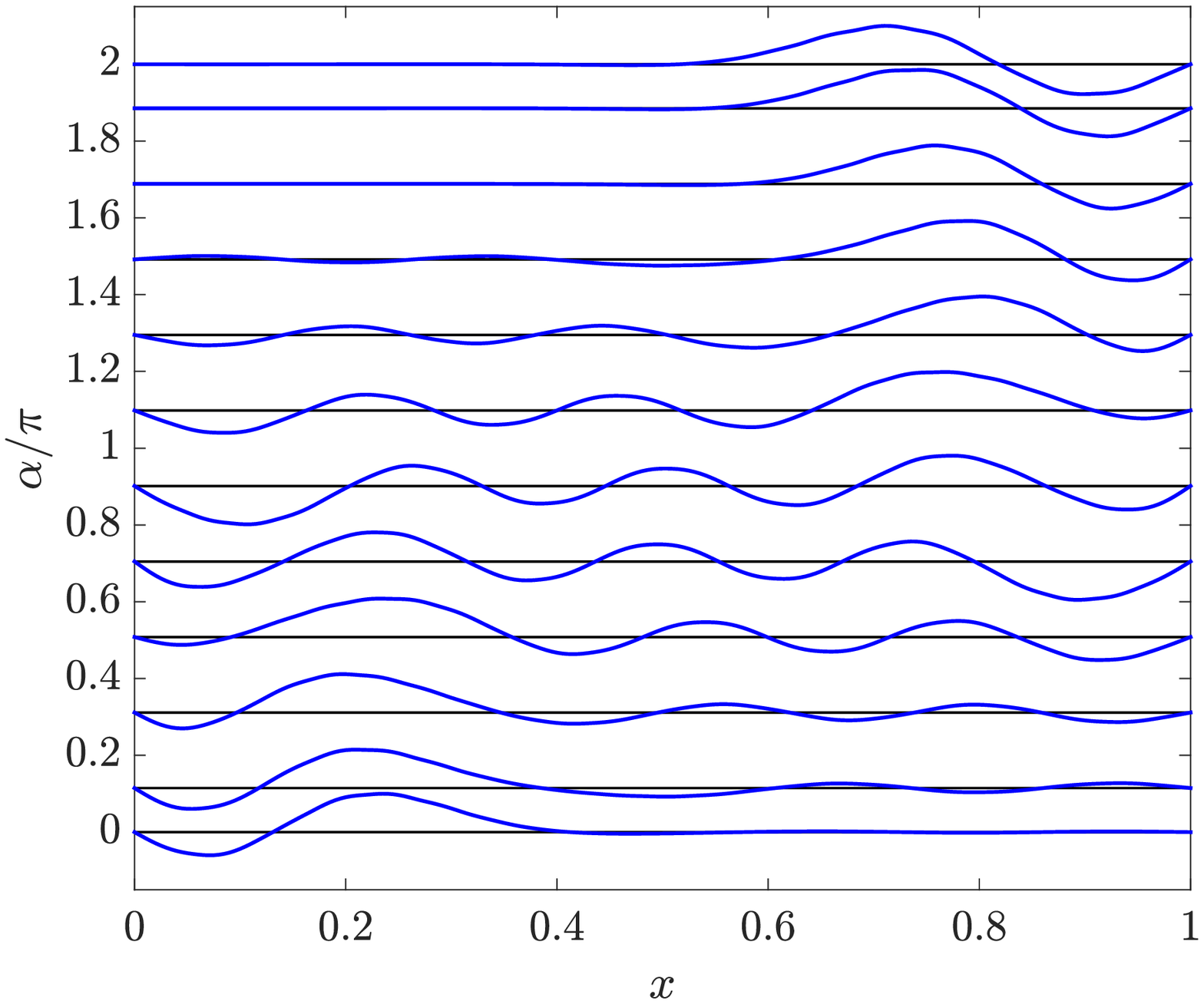}
	}
	\caption{Mode transitions for beam with $\theta=0.06$, $S_f=20$ and $\gamma_g=10$. The normalized mode shapes are displayed as a function of $\alpha$ for the branches tracked in the spectrum of Fig.~\ref{spectrumphase}. The set of modes displayed in (a-d) correspond to the red dots marked in the (a-d) branches of Fig.~\ref{spectrumphase}.} 
	\label{modesphase}
\end{figure}

Herein, we show that transitions of this type are generally related to transitions of the edge states as they span the gaps and merge with the bulk as a function of the phase $\alpha$. To illustrate, we compute how the bulk and finite beam spectra of Fig.~\ref{butterflyzoom} varies with $\alpha$ for fixed $\theta=0.06$, which is displayed in Fig.~\ref{spectrumphase}. The black dots correspond to bulk modes, while red dots correspond to modes of the finite beam with $S_f=20$. The computation of the spectrum reveals that edge modes span the gaps also for variations of $\alpha$, and that these modes undergo transitions when they merge with the bulk bands. A few selected modes of the finite beam are tracked and marked by thick red lines, and the variation of their mode shapes with $\alpha$ is displayed in Fig.~\ref{modesphase}. For $\alpha=0$, the (a) mode is the same as previously shown (Fig.~\ref{modetransition1}), right after the edge mode branch marked in Fig.~\ref{butterflyzoom} merged with the bulk, and is localized at the right boundary. As $\alpha$ varies, its branch remains in the bulk and its mode shape translates within the bulk (Fig.~\ref{modephase1}), and eventually it localizes at the left boundary when the branch detaches from the bulk. Note that the location of the ground springs is $2\pi-$periodic with $\alpha$, \textit{i.e.} $x_s(\alpha+2\pi)=x_s(\alpha)$, which results in the same periodicity of the eigenfrequencies of the spectrum, \textit{i.e.} $\Omega(\alpha+2\pi)=\Omega(\alpha)$. After tracking the mode branches, one can observe that the (a) branch is connected to the (b) branch, while the (c) branch is connected to the (d) branch. For example, the eigenfrequency of the (b) branch for $\alpha=0$ is equal to the eigenfrequency of the (a) branch for $\alpha=2\pi$, and corresponds to a vibration mode localized at the left boundary. As $\alpha$ varies, the (b) mode branch transverses the gap as a left-localized mode, touches the bulk and becomes a bulk mode. It then continues to span the gaps as a mode localized at the right boundary (Fig.~\ref{modephase2}). At $\alpha=2\pi$, the (b) mode reaches the same point as (a) mode for $\alpha=0$, thus completing a cycle. A similar cycle is illustrated for the (c) and (d) modes, which have their mode shapes displayed in Figs.~\ref{modesphase}(c,d). 

\begin{figure}[b!]
	\centering
\subfigure[]{\label{spectrumphase2}
		\includegraphics[width=0.475\textwidth]{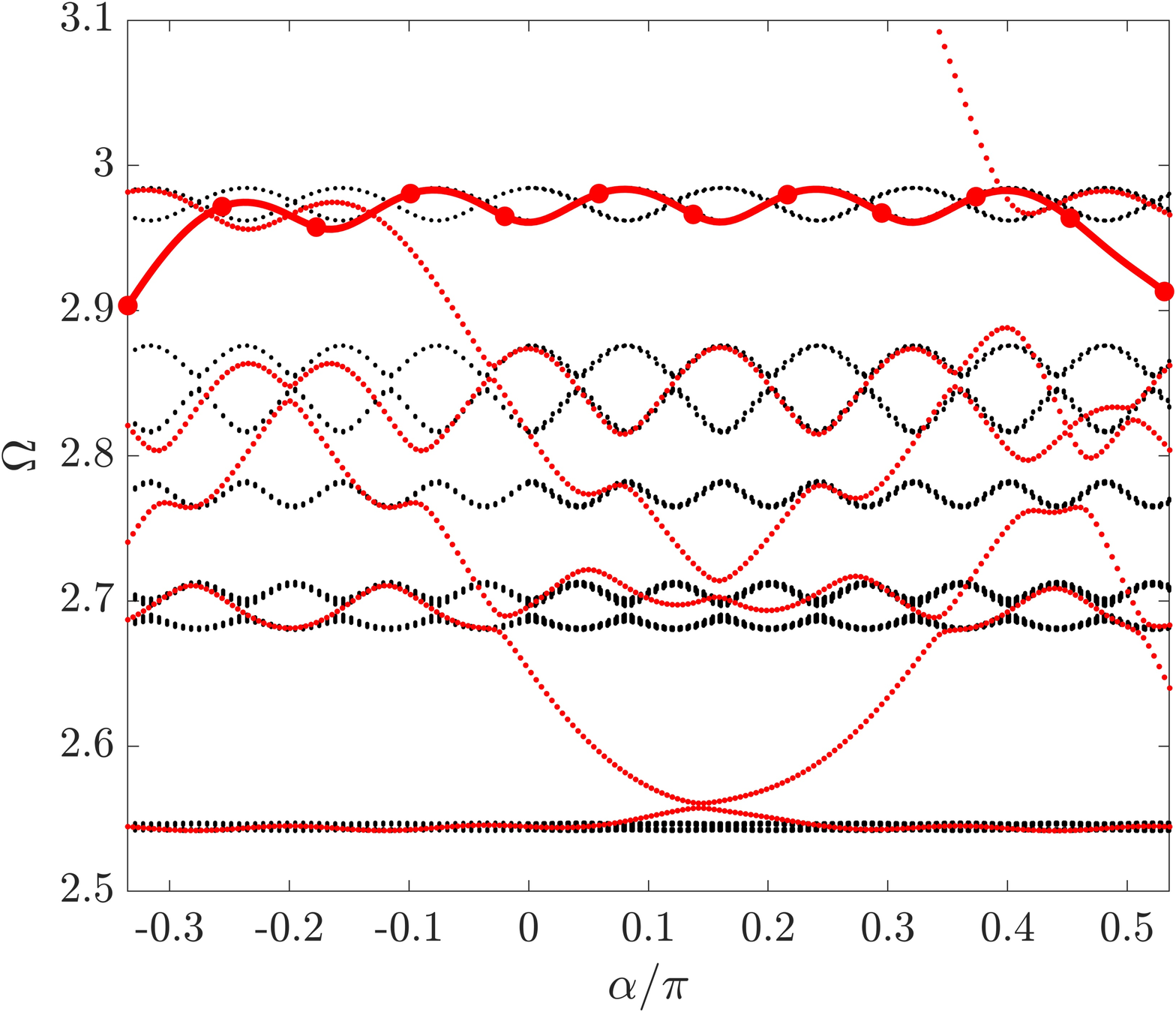}
	}
	\subfigure[]{\label{modephase5}
		\includegraphics[width=0.475\textwidth]{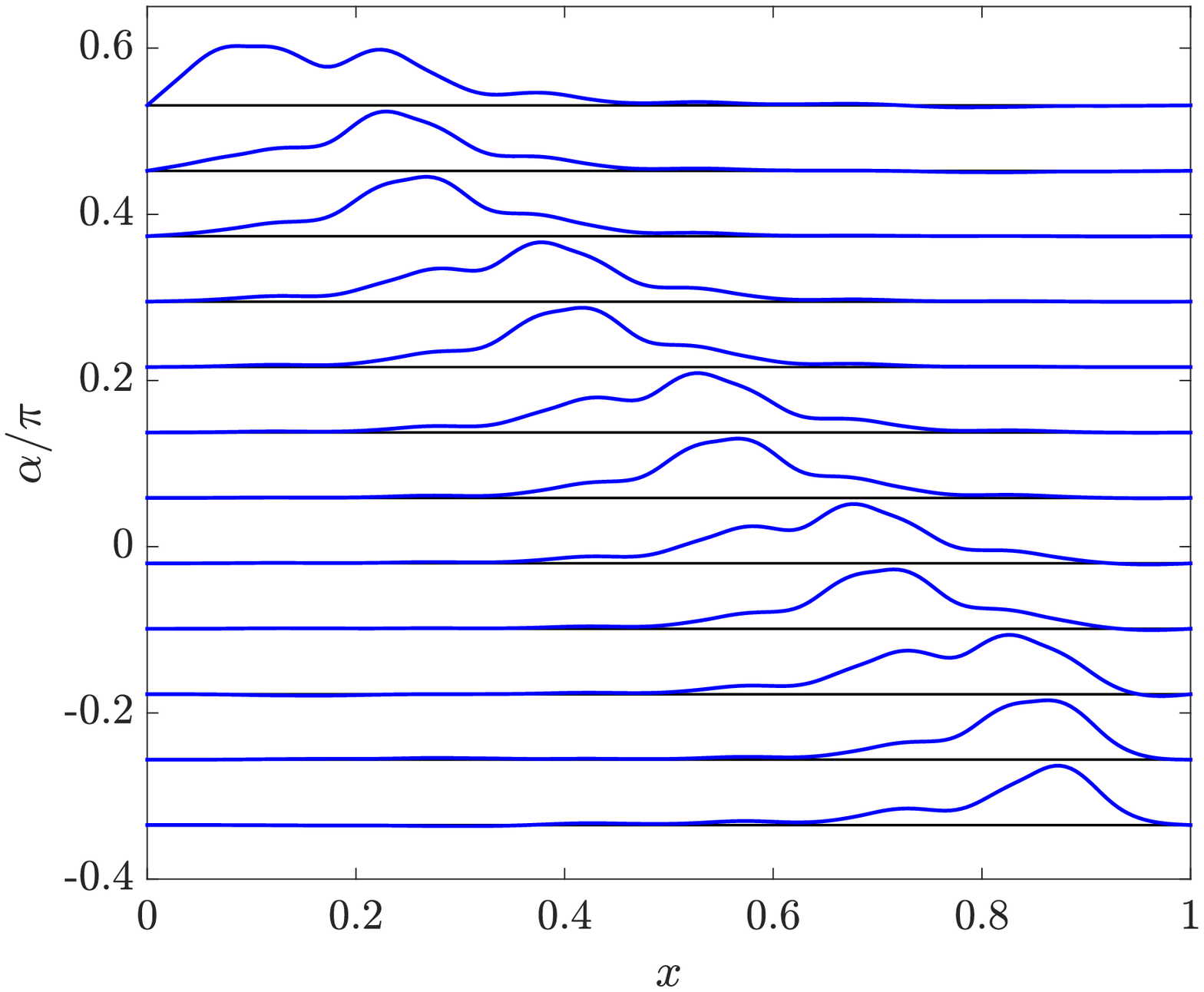}
	}
	\caption{Spectrum as a function of $\alpha$ (a) and mode transition of the branch marked by the thick red line (b) for beam with $\theta=0.36$, $S_f=20$ and $\gamma_g=10$. } 
	\label{FigInterior3}
\end{figure}

Therefore, the localized modes in the interior of the domain arise as topological states merge with the bulk and change their localization edge as a function of $\alpha$. As another example, we consider the mode branch tracked in the spectrum of Fig.~\ref{butterflyzoom2}. For $\theta=0.36$, the mode shape is localized in a region inside the domain of the beam (Fig.~\ref{modetransition2}). By computing the spectrum for $\theta=0.36$ as a function of $\alpha$ (Fig.~\ref{spectrumphase2}), one can see that the vibration mode localized in the interior (for $\alpha=0$) arises as a result of the transition of the edge state merging with the bulk and changing from a localized state at the right boundary to a localized state at the left boundary (Fig.~\ref{modephase5}). 

We emphasize that the modes localized in the interior of the domain are in fact bulk modes, and that they are not localized for structures of any size. In some cases, increasing the size of the structure may transform these modes into globally spanning modes where the localization regions repeat in regular intervals. In the literature of discrete QP lattices, this type of localized modes have been investigated in the context of phase transitions~\cite{aubry1980analyticity,lahini2009observation,martinez2018quasiperiodic}, but a connection to the edge states spanning the gaps as given here is currently missing. This connection identifies an open question regarding the regions where these modes may localize, which may be explained in the context of the smooth edge-interior-edge transitions experienced by some of the topological states as a function of $\alpha$. For some modes, like the one displayed in Fig.~\ref{modephase1}, this transition occurs while maintaining the mode shape essentially unaltered, in a manner that is vaguely reminiscent of solitons~\cite{dauxois2006physics}. These transitions may be exploited to produce novel physical behavior like adiabatic pumping through a second dimension through a continuous translation of the localized mode, in contrast to the topological pumps realized so far~\cite{kraus2012topological,rosa2018edge} that rely on edge-bulk-edge transitions like the one shown in Fig.~\ref{modephase2}.

\section{Conclusions}\label{concSec}
We investigate beams with QP arrays of ground springs whose locations are obtained by projecting from periodic arrays of circles. A family of beams is generated by varying a parameter of the projection and the fundamental properties of this class of structures are investigated through numerical simulations. A fractal nature of the bulk spectrum is observed by varying the QP parameter, and its topological properties are described in terms of the integrated density of states. We then illustrate how topological modes that are localized at a boundary span the gaps as the projection parameter is varied, and how the number of topological modes is given by topological invariants related to the integrated density of states. Finally, we  illustrate mode transitions whereby topological edge modes migrate and localize in the interior of the structure as their branches merges with the bulk bands as a function of one of the projection parameters. The presented results suggest avenues for systematically designing elastic structures that support localized modes at desired locations by exploiting disorder in the form of deterministic patterns. While ground springs were used here, similar studies may be conducted in structures where added mass or resonators may be employed. The mode transitions demonstrated herein can be further explored to produce novel topological pumps in a soliton-like manner, that is, a localized mode that slowly travels with unaltered shape. Future work may focus on extending the results presented here to other classes of 1D QP elastic media, on experimental demonstrations, and on 2D QP elastic media.

\section*{Acknowledgments}
The authors gratefully acknowledge funding support from the National Science Foundation through the CMMI 1719728 and EFRI 1741685 grants.

\bibliographystyle{unsrt}
\bibliography{paper}

\begin{thebibliography}{10}

\bibitem{anderson1958absence}
PW~Anderson.
\newblock Absence of diffusion in certain random lattices.
\newblock {\em Physical review}, 109(5):1492, 1958.

\bibitem{hu2008localization}
H~Hu, A~Strybulevych, JH~Page, SE~Skipetrov, and BA~van Tiggelen.
\newblock Localization of ultrasound in a three-dimensional elastic network.
\newblock {\em Nature Physics}, 4(12):945, 2008.

\bibitem{sheng1990scattering}
P~Sheng.
\newblock {\em Scattering and localization of classical waves in random media},
  volume~8.
\newblock World Scientific, 1990.

\bibitem{han2008wave}
P~Han, CT~Chan, and ZQ~Zhang.
\newblock Wave localization in one-dimensional random structures composed of
  single-negative metamaterials.
\newblock {\em Physical Review B}, 77(11):115332, 2008.

\bibitem{sievers1988intrinsic}
AJ~Sievers and S~Takeno.
\newblock Intrinsic localized modes in anharmonic crystals.
\newblock {\em Physical Review Letters}, 61(8):970, 1988.

\bibitem{campbell2004localizing}
DK~Campbell, S~Flach, YS~Kivshar, et~al.
\newblock Localizing energy through nonlinearity and discreteness.
\newblock {\em Physics Today}, 57(1):43--49, 2004.

\bibitem{page1990asymptotic}
JB~Page.
\newblock Asymptotic solutions for localized vibrational modes in strongly
  anharmonic periodic systems.
\newblock {\em Physical Review B}, 41(11):7835, 1990.

\bibitem{fishman2012nonlinear}
S~Fishman, Y~Krivolapov, and A~Soffer.
\newblock The nonlinear schr{\"o}dinger equation with a random potential:
  results and puzzles.
\newblock {\em Nonlinearity}, 25(4):R53, 2012.

\bibitem{janot2012quasicrystals}
C~Janot.
\newblock {\em Quasicrystals: A Primer}.
\newblock OUP Oxford, 2012.

\bibitem{steinhardt1987physics}
PJ~Steinhardt and S~Ostlund.
\newblock {\em The physics of quasicrystals}.
\newblock World Scientific, 1987.

\bibitem{man2005experimental}
W~Man, M~Megens, PJ~Steinhardt, and PM~Chaikin.
\newblock Experimental measurement of the photonic properties of icosahedral
  quasicrystals.
\newblock {\em Nature}, 436(7053):993, 2005.

\bibitem{liu2004governing}
GT~Liu, TY~Fan, and RP~Guo.
\newblock Governing equations and general solutions of plane elasticity of
  one-dimensional quasicrystals.
\newblock {\em International journal of solids and structures},
  41(14):3949--3959, 2004.

\bibitem{hofstadter1976energy}
DR~Hofstadter.
\newblock Energy levels and wave functions of bloch electrons in rational and
  irrational magnetic fields.
\newblock {\em Physical review B}, 14(6):2239, 1976.

\bibitem{aubry1980analyticity}
S~Aubry and G~Andr{\'e}.
\newblock Analyticity breaking and anderson localization in incommensurate
  lattices.
\newblock {\em Ann. Israel Phys. Soc}, 3(133):18, 1980.

\bibitem{lahini2009observation}
Yoav Lahini, Rami Pugatch, Francesca Pozzi, Marc Sorel, Roberto Morandotti, Nir
  Davidson, and Yaron Silberberg.
\newblock Observation of a localization transition in quasiperiodic photonic
  lattices.
\newblock {\em Physical review letters}, 103(1):013901, 2009.

\bibitem{martinez2018quasiperiodic}
Alejandro~J Mart{\'\i}nez, Mason~A Porter, and PG~Kevrekidis.
\newblock Quasiperiodic granular chains and hofstadter butterflies.
\newblock {\em arXiv preprint arXiv:1801.09860}, 2018.

\bibitem{segev2013anderson}
M~Segev, Y~Silberberg, and DN~Christodoulides.
\newblock Anderson localization of light.
\newblock {\em Nature Photonics}, 7(3):197, 2013.

\bibitem{zhu2018simultaneous}
W~Zhu, X~Fang, D~Li, Y~Sun, Y~Li, Y~Jing, and H~Chen.
\newblock Simultaneous observation of topological edge state and exceptional
  point in an open and non-hermitian system.
\newblock {\em arXiv preprint arXiv:1803.04110}, 2018.

\bibitem{kraus2016quasiperiodicity}
Yaacov~E Kraus and Oded Zilberberg.
\newblock Quasiperiodicity and topology transcend dimensions.
\newblock {\em Nature Physics}, 12(7):624, 2016.

\bibitem{ozawa2016synthetic}
Tomoki Ozawa, Hannah~M Price, Nathan Goldman, Oded Zilberberg, and Iacopo
  Carusotto.
\newblock Synthetic dimensions in integrated photonics: From optical isolation
  to four-dimensional quantum hall physics.
\newblock {\em Physical Review A}, 93(4):043827, 2016.

\bibitem{kitaev2009periodic}
A~Kitaev.
\newblock Periodic table for topological insulators and superconductors.
\newblock In {\em AIP Conference Proceedings}, volume 1134, pages 22--30. AIP,
  2009.

\bibitem{kraus2012topological}
YE~Kraus, Y~Lahini, Z~Ringel, M~Verbin, and O~Zilberberg.
\newblock Topological states and adiabatic pumping in quasicrystals.
\newblock {\em Physical review letters}, 109(10):106402, 2012.

\bibitem{verbin2013observation}
M~Verbin, O~Zilberberg, YE~Kraus, Y~Lahini, and Y~Silberberg.
\newblock Observation of topological phase transitions in photonic
  quasicrystals.
\newblock {\em Physical review letters}, 110(7):076403, 2013.

\bibitem{zilberberg2018photonic}
Oded Zilberberg, Sheng Huang, Jonathan Guglielmon, Mohan Wang, Kevin~P Chen,
  Yaacov~E Kraus, and Mikael~C Rechtsman.
\newblock Photonic topological boundary pumping as a probe of 4d quantum hall
  physics.
\newblock {\em Nature}, 553(7686):59, 2018.

\bibitem{apigo2018topological}
David~J Apigo, Kai Qian, Camelia Prodan, and Emil Prodan.
\newblock Topological edge modes by smart patterning.
\newblock {\em Physical Review Materials}, 2(12):124203, 2018.

\bibitem{apigo2019observation}
David~J Apigo, Wenting Cheng, Kyle~F Dobiszewski, Emil Prodan, and Camelia
  Prodan.
\newblock Observation of topological edge modes in a quasiperiodic acoustic
  waveguide.
\newblock {\em Physical Review Letters}, 122(9):095501, 2019.

\bibitem{gei2010wave}
M~Gei.
\newblock Wave propagation in quasiperiodic structures: stop/pass band
  distribution and prestress effects.
\newblock {\em International Journal of Solids and Structures},
  47(22-23):3067--3075, 2010.

\bibitem{morini2018waves}
Lorenzo Morini and Massimiliano Gei.
\newblock Waves in one-dimensional quasicrystalline structures: dynamical trace
  mapping, scaling and self-similarity of the spectrum.
\newblock {\em Journal of the Mechanics and Physics of Solids}, 2018.

\bibitem{chen2008elastic}
A~Chen, Y~Wang, G~Yu, Y~Guo, and Z~Wang.
\newblock Elastic wave localization in two-dimensional phononic crystals with
  one-dimensional quasi-periodicity and random disorder.
\newblock {\em Acta Mechanica Solida Sinica}, 21(6):517--528, 2008.

\bibitem{meirovitch1975elements}
L~Meirovitch.
\newblock {\em Elements of vibration analysis}.
\newblock McGraw-Hill, 1975.

\bibitem{ern2013theory}
A~Ern and J-L Guermond.
\newblock {\em Theory and practice of finite elements}, volume 159.
\newblock Springer Science \& Business Media, 2013.

\bibitem{bellissard1986k}
Jean Bellissard.
\newblock K-theory of c*—algebras in solid state physics.
\newblock In {\em Statistical mechanics and field theory: mathematical
  aspects}, pages 99--156. Springer, 1986.

\bibitem{prodan2018k}
E~Prodan and Y~Shmalo.
\newblock The k-theoretic bulk-boundary principle for dynamically patterned
  resonators.
\newblock {\em arXiv preprint arXiv:1805.10629}, 2018.

\bibitem{prodan2016bulk}
E~Prodan and H~Schulz-Baldes.
\newblock Bulk and boundary invariants for complex topological insulators.
\newblock {\em K}, 2016.

\bibitem{dauxois2006physics}
Thierry Dauxois and Michel Peyrard.
\newblock {\em Physics of solitons}.
\newblock Cambridge University Press, 2006.

\bibitem{rosa2018edge}
Matheus~IN Rosa, Raj~Kumar Pal, Jos{\'e}~RF Arruda, and Massimo Ruzzene.
\newblock Edge states and topological pumping in elastic lattices with
  periodically modulated coupling.
\newblock {\em arXiv preprint arXiv:1811.02637}, 2018.

\end{thebibliography}

\section*{Appendix: Spectral flow between bulk bands: an illustrative example}
To elucidate the concept of spectral flow between bulk bands which manifest as localized modes at the boundary of QP beams, we here consider simple spring-mass chains consisting of point masses connected by linear springs. The stiffness $k_n$ of springs connecting masses $n$ and $n+1$ varies in space following the same modulation rule as Eqn.~\eqref{eq: beamQP}: $k_n = 1 + \lambda \sin (2 \pi \theta n)$, while the masses are all identical with value $m=1$. We consider a finite chain of $N=40$ masses with the first and last mass connect by a spring $k_{40}$ so that the chain mathematically resembles a ring. We examine how the eigenvalues (natural frequencies) of this chain change as we vary $\theta$ in the range $[1/5,1/4]$. 

\begin{figure}[h!]
\centering
\subfigure[]{
\includegraphics[width=0.42\textwidth]{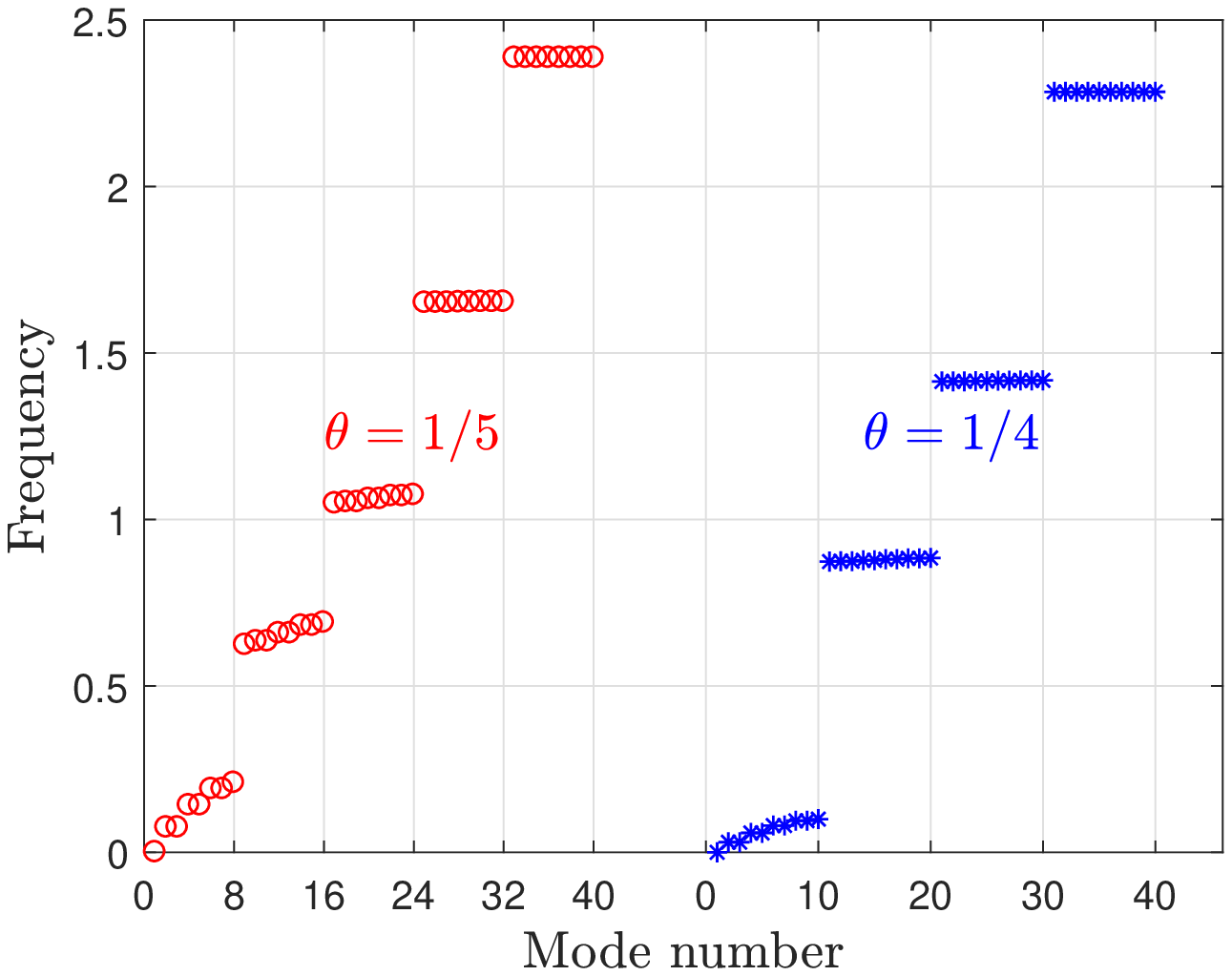}\label{SpringMass_1}
}
\subfigure[]{
\includegraphics[width=0.5\textwidth]{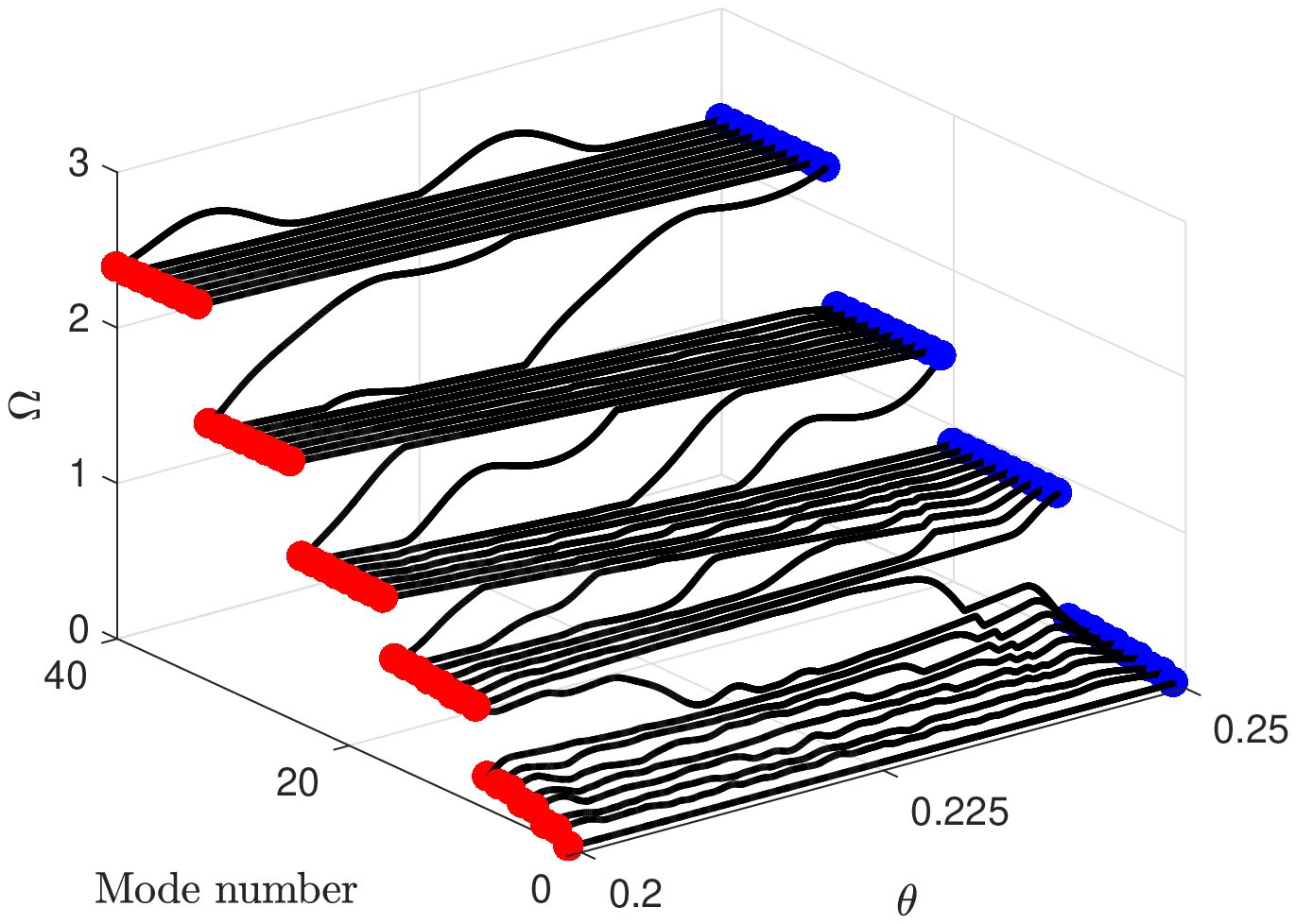}\label{SpringMass_2}
}
\label{SpringMassFig}
\caption{(a) Natural frequencies of a spring mass chain with $\theta=1/5$ (left) and $\theta=1/4$ (right) showing a change in the number of bands and the number of modes in each band with $\theta$. (b) Change in natural frequencies with $\theta$ showing migration of spectra from one bulk band to another.} 
\end{figure}

Let us first analyze the chain with stiffness modulation according to $\theta=1/5$. This chain is periodic with $5$ masses per unit cell and its dispersion diagram has $5$ branches. 
Note that the periodicity condition implies that all the natural frequencies of this chain are given by the summation of Bloch eigenstates with wavenumbers $\{-\mu,\mu\}$, with $\mu$ taking values $2\pi s/5$, $s\in \{1,2,...,5\}$. The natural frequencies lie on its dispersion branches at locations where the wavenumbers are  $2\pi s/5,\;s\in\{1,2,...,5\}$. Hence, the $40$ natural frequencies are distributed equally in the $5$ branches, with $8$ modes in each branch. Next, let us consider a chain with the quasiperiodic parameter set to $\theta=1/4$. Analogous to the previous case, we now have the  $40$ modes distributed equally among the $4$ dispersion branches with $10$ modes in each branch. Both these cases are illustrated in Fig.~\ref{SpringMass_1} and we observe that the distribution of natural frequencies is consistent with the description presented above. 

Now, keeping the chain length fixed ($S=40$), let us vary $\theta$ in the range $[1/5,1/4]$ and analyze how the frequencies change. Figure~\ref{SpringMass_2} displays how these $40$ natural frequencies change with $\theta$. Let us consider the first bulk band for $\theta = 1/5=0.2$ and $1/4=0.25$. The number of modes change from $8$ to $10$ as $\theta$ changes in this range. We note that the change in an eigenvalue with $\theta$ is a continuous function. This continuity requirement along with the requirement of a change in the number of modes in each bulk band as $\theta$ changes from $0.2$ to $0.25$ results in a \textit{spectral flow}, i.e., eigenvalues migrate from other bulk bands. These requirements are termed as topological constraints since they arise from the continuity of the eigenvalue and from the periodicity conditions on a ring. Indeed, we observe in Fig.~\ref{SpringMass_2} that two modes migrate to the first band from the band above it to satisfy the condition of the number of modes increasing from $8$ to $10$.  

A similar migration of modes is observed in the top band, where two modes migrate from the band below (at $\theta=0.2$) resulting in the number of modes increasing from $8$ (at $\theta=0.2$) to $10$ at ($\theta=0.25$). Next, let us consider the band below the top band. It also has $8$ modes at $\theta = 0.2$ and $10$ modes at $\theta=0.25$. As $\theta$ changes, we observe that the remaining $6$ modes of this band at $\theta=0.2$ move to become the top $6$ modes at $\theta=0.25$, while $4$ modes from the band below move to satisfy the topological constraint of $10$ modes at $\theta = 0.2$. Similar considerations show that $6$ modes migrate up between the two second bands as $\theta$ changes from $0.2$ to $0.25$, resulting in the number of modes changing from $8$ to $10$.  We have thus shown how modes migrate between bulk bands to satisfy topological constraints resulting from periodicity. All these modes span the bandgap as they migrate from one bulk band to another. In the finite continuous structures that we consider in this work, there is a similar change in the number of modes  and these migrating modes manifest as localized modes at the boundary. 

\end{document}